\documentclass[prl,twocolumn,10pt,aps,longbibliography]{revtex4-1} %superscriptaddress
\usepackage{amsmath,amssymb,bm}
\usepackage{graphicx}
\usepackage{epstopdf}
\usepackage{latexsym}
\usepackage{subfigure}
\usepackage{color}
\usepackage{natbib}
\usepackage{hyperref}
\usepackage{braket}
\usepackage{dsfont}
\usepackage[english]{babel}
\usepackage{blindtext}
\hypersetup{
  colorlinks,
  citecolor=magenta,
  linkcolor=blue,
  urlcolor=blue}
      
\begin{document}

\title{Lee-Yang zeros at $O(3)$ and deconfined quantum critical points}
\author{Jonathan D'Emidio}
\email{jonathan.demidio@dipc.org}
\affiliation{Donostia International Physics Center, P. Manuel de Lardizabal 4, 20018 Donostia-San Sebasti\'an, Spain}

\begin{abstract}
Lee-Yang theory, based on the study of zeros of the partition function, is widely regarded as a powerful and complimentary approach to the study of critical phenomena and forms a foundational part of the theory of phase transitions.  Its widespread use, however, is complicated by the fact that it requires introducing complex-valued fields that create an obstacle for many numerical methods, especially in the quantum case where very limited studies exist beyond one dimension.  Here we present a simple and statistically exact method to compute partition function zeros with general complex-valued external fields in the context of large-scale quantum Monte Carlo simulations.  We demonstrate the power of this approach by extracting critical exponents from the leading Lee-Yang zeros of 2D quantum antiferromagnets with a complex staggered field, focusing on the Heisenberg bilayer and square-lattice $J$-$Q$ models.  The method also allows us to introduce a complex field that couples to valence bond solid order, where we observe extended rings of zeros in the $J$-$Q$ model with purely imaginary staggered and valence bond solid fields.
\end{abstract}
\maketitle

Introduced over a half century ago, the Lee-Yang theory of phase transitions [\onlinecite{YangLee1952:StatTheoryI},\onlinecite{LeeYang1952:StatTheoryII}] provides deep insights into how non-analyticities of the free energy arise in the thermodynamic limit of finite-size systems.  The theory rests on the fact that while finite systems are manifestly analytic in the domain of real-valued physical control parameters, they can display singularities---zeros of the partition function---when the control parameters are extended to the complex plane.  As the thermodynamic limit is approached, these zeros can accumulate near and eventually pinch the real axis, producing a genuine phase transition.  In fact, even when zeros accumulate away from the real-axis they provide an interesting example of non-unitary critical points [\onlinecite{Fisher1978:LYEphi3}, \onlinecite{Cardy1985:ConformalLYE}].

While the study of Lee-Yang zeros may seem relegated to the realm of pure theory, they have in fact been demonstrated in several experimental systems [\onlinecite{Binek1998:LYZmag,Peng2015:ExpLYZ,Brandner2017:ExpDynLYZ}] and quantum computers [\onlinecite{Akhil2021:QCLYZ}].  But perhaps the greatest impact of the theory lies in computational studies of many-body systems where the themes range from protein folding [\onlinecite{Lee2013:LYZprotein}] and DNA zippers [\onlinecite{Deger2018:LYZzipper},\onlinecite{Majumdar2020:LYZcumulant}] to quantum chromodynamics [\onlinecite{Nagata2015:LYZQCD}], to name but a few. See Ref. [\onlinecite{Bena2005:YangLeeReview}] for a review.  However, despite the broad range of studies, there remains a glaring disparity between the strength of numerical methods in the classical versus the quantum case, an issue that we seek to address in this work.

While classical systems with complex fields can be efficiently treated with a variety of methods, from histogram-based approaches [\onlinecite{Ferrenberg1988:MCtransition, Kenna1993:LYZlogphi4, Kenna1994:LYZ4dIsing}] or high-order cumulants [\onlinecite{Flindt2013:TrajPhaseTransLYZ},\onlinecite{Deger2020:LYZcumulantIsing}] to tensor network methods [\onlinecite{Shimizu2014:GrassmanTNSchwinger,Shimizu2014:GrassmanTNCriticalSchwinger,GarciaSaez2015:LYZTN,Hong2022:TNLYZXY}], the quantum case is more restrictive, mainly being limited to 1D [\onlinecite{Heyl2013:DynamicalQPT,Ananikian2014:Diamond,Peotta2021:DynamicalQPT}] or small 2D lattices [\onlinecite{Kist2021:LYTheoryQMB,Vecsei2022:LYZ2DquantumIsing,Brange2022:Dynamical2Dquantum,vecsei2023leeyang}].  Here we develop a statistically exact method for extracting Lee-Yang zeros in large-scale quantum Monte Carlo (QMC) simulations of spin systems.  The method builds on a simple, yet so far unused, formula for computing free energy differences in QMC, and allows for the computation of partition function ratios for a whole range of complex external fields based a single simulation in zero field.

%\begin{equation}
%\label{eq:sse}
%Z(\beta)= \text{Tr}\left(e^{-\beta H}\right) = \sum^{\infty}_{n=0} \frac{\beta^n}{n!} \text{Tr}\left( (-H)^n \right).
%\end{equation}

{\em Free energy differences in QMC:} We work in the context of the stochastic series expansion (SSE) quantum Monte Carlo algorithm [\onlinecite{Sandvik2010:CompStud}], which is a statistically exact method for sampling the partition function as a Taylor expansion: $Z(\beta)= \text{Tr}\left(e^{-\beta H}\right) = \textstyle\sum^{\infty}_{n=0} \frac{\beta^n}{n!} \text{Tr}\left( (-H)^n \right).$  The trace is then further decomposed into a sum of powers of Hamiltonian matrix elements by inserting complete sets of states, giving a sum over SSE configurations [\onlinecite{Sandvik2010:CompStud}].

In this formulation it is simple to show that partition function ratios at different inverse temperatures can be computed as follows:
\begin{equation}
\label{eq:zrat}
R_{\beta}(\tilde{\beta}) \equiv \frac{Z(\tilde{\beta})}{Z(\beta)}= \left\langle  \left(\frac{\tilde{\beta}}{\beta}\right)^n \right\rangle_{\beta},
\end{equation}
where $n$ is the expansion order of the SSE configuration and the average is taken in the ensemble with inverse temperature $\beta$.  To the best of our knowledge, Eqn. (\ref{eq:zrat}) has not yet been used nor has appeared in the literature. 

The ratio formula can be extended to the case where the partition functions differ only in the value of a Hamiltonian coupling $J$.  In this case we simply replace $\beta \to J$, $\tilde{\beta} \to \tilde{J}$, and $n \to n_{J}$ in Eq. (\ref{eq:zrat}) where $n_J$ refers to the number of $J$-type matrix elements in the SSE configuration.

Note that in these computations, only the denominator partition function needs to be simulated, and the ratio with any ``nearby" partition function can be computed by recording a histogram of the number of operators ($n$) during an SSE simulation.  This formula then allows us to extend the ratio estimator to complex values of its argument, which is the focus of this work.  Before demonstrating this, we first will describe how to introduce a complex-valued staggered field that couples to the N\'eel order parameter, which will allow for the meaningful extraction of Lee-Yang zeros in the case of quantum anti-ferromagnets.

\begin{figure}[!t]
\centerline{\includegraphics[angle=0,width=0.95\columnwidth]{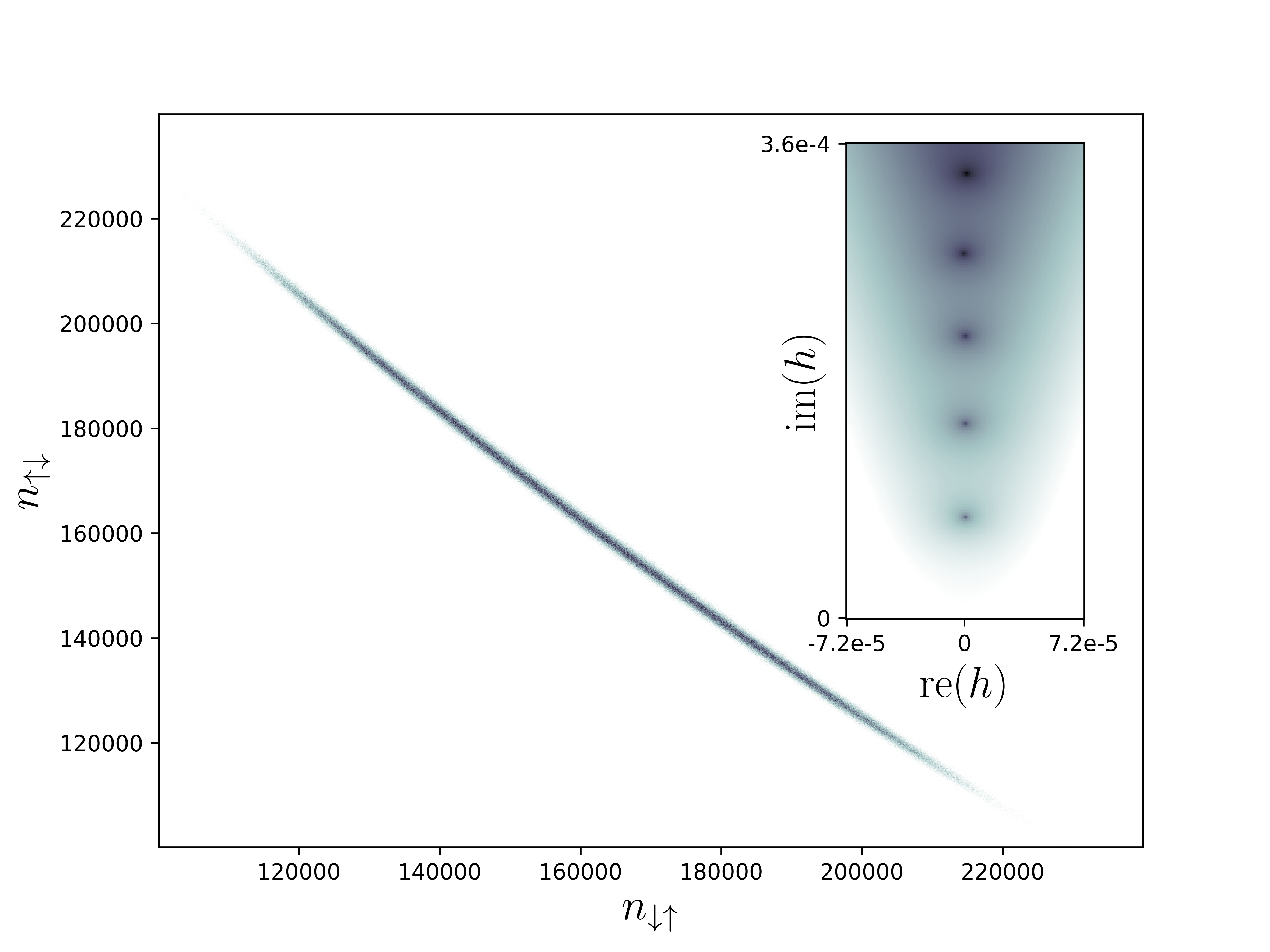}}
\caption{Main panel: A histogram of the number operators from Eq. (\ref{eq:zrath}) for the Heisenberg bilayer with $L=80$, $J_{\perp}=2.5222$ and $\beta=40$ with $J=1$.  Inset:  Eq. (\ref{eq:zrath}) applied to the histogram with complex-valued external N\'eel field $h$, where we plot $-\ln(|R_0(h)|)$ and the Lee-Yang zeros appear as the dark points.}
\label{fig:bilayerhisto}
\end{figure}

{\em Lee-Yang fields for quantum antiferromagnets:}
The classic case for Lee-Yang zeros is that of the Ising model in an external magnetic field, which is taken to be complex.  The Lee-Yang circle theorem [\onlinecite{LeeYang1952:StatTheoryII}] then states that all of the Ising partition function zeros with a complex magnetic field lie on the imaginary axis.  In order to extend this picture to quantum antiferromagnets we introduce a field that couples to the N\'eel order parameter.  Here, remarkably, we will show that it is possible to compute the ratio at finite complex field while simulating the partition function in zero field.

The idea is to \textit{imagine} embedding the $z-$component of the N\'eel field operator inside nearest-neighbor operators that are already present in the Hamiltonian $h_{i,j} = J(\vec{S}_i \cdot \vec{S}_j -\tfrac{1}{4}) \to J(\vec{S}_i \cdot \vec{S}_j -\tfrac{1}{4})  +\tfrac{h}{N_c}(S^z_i - S^z_j)$.  Here $i\in$ sublattice $A$ and $j\in$ sublattice $B$ and $N_c$ is the number of $J$-operators that touch each site (usually the coordination number of the lattice).  The diagonal matrix elements then change as $-\langle \uparrow_i \downarrow_j | h_{i,j} | \uparrow_i \downarrow_j  \rangle = \tfrac{J}{2} \to  \tfrac{J}{2} - \tfrac{h}{N_c}$ and $-\langle \downarrow_i \uparrow_j | h_{i,j} | \downarrow_i \uparrow_j  \rangle = \tfrac{J}{2} \to  \tfrac{J}{2} + \tfrac{h}{N_c}$.  The ratio of these matrix elements before and after the change gives $1 \mp \frac{2 h}{N_c J}$.  If we denote the number of such matrix elements in an SSE configuration by $n_{\uparrow\downarrow}$ and $n_{\downarrow\uparrow}$ then the ratio formula with a N\'eel field reads
\begin{equation}
\label{eq:zrath}
R_{0}(h)= \left\langle  \left(1-\frac{2 h}{N_c J}\right)^{n_{\uparrow\downarrow}}  \left(1+\frac{2 h}{N_c J}\right)^{n_{\downarrow\uparrow}} \right\rangle_{h=0}.
\end{equation}
We emphasize that this formula only requires a histogram of the values $(n_{\uparrow\downarrow},n_{\downarrow\uparrow})$ obtained by QMC simulation (see Fig. (\ref{fig:bilayerhisto})) of typical Heisenberg models in the absence of external fields, and allows for the statistically exact extraction of partition function zeros with a complex-valued N\'eel field.  We now demonstrate this with a concrete example, the $O(3)$ quantum critical Heisenberg bilayer.

{\em Heisenberg bilayer:} As a first test case for this technique we select a well studied 2D quantum spin model for the $O(3)$ transition in $d=2+1$ dimensions, the Heisenberg bilayer given by the Hamiltonian
\begin{equation}
\label{eq:Hbl}
H_{\text{bl}}= J\sum_{\langle i,j \rangle, a=1,2}(\vec{S}_{i,a} \cdot \vec{S}_{j,a} -\tfrac{1}{4}) + J_{\perp}\sum_{i}(\vec{S}_{i,1} \cdot \vec{S}_{i,2} -\tfrac{1}{4}),
\end{equation}
where $\langle i,j \rangle$ are nearest neighbor pairs of a square lattice and $a$ is the layer index.  When $J_{\perp}=0$ the model reduces to two independent Heisenberg models, which exhibits long-range N\'eel order at zero temperature, while for strong $J_{\perp}$ singlets are formed on the interlayer bonds and N\'eel order is destroyed.  A continuous transition in the $O(3)$ universality class occurs at $J_{\perp}/J=2.5220(1)$ [\onlinecite{Wang2006:Bilayer}].

Here we will focus on extracting the Lee-Yang zeros, zeros of Eq. (\ref{eq:zrath}) at the critical point.  First we show that the zeros of the partition function with a complex-valued N\'eel field lie exactly on the imaginary axis, akin to the Lee-Yang circle theorem for the Ising model [\onlinecite{LeeYang1952:StatTheoryII}] and its generalization to the classical Heisenberg model [\onlinecite{Harris1970:GenLeeYang}].  This is shown if Fig. (\ref{fig:bilayerhisto}), where in the main panel we display the histogram of the number operators from Eq. (\ref{eq:zrath}) that are used to compute the partition function ratio.  The data is collected for a bilayer system with side length $L=80$ at the critical point, fixing $J=1,J_{\perp}=2.5222$ and $\beta=L/2$.  The inset shows $-\ln(|R_0(h)|)$ computed with this histogram, where the dark spikes indicate the first five zeros near $|h|=0$, which lie directly on the imaginary axis. 

\begin{figure}[!t]
\centerline{\includegraphics[angle=0,width=0.95\columnwidth]{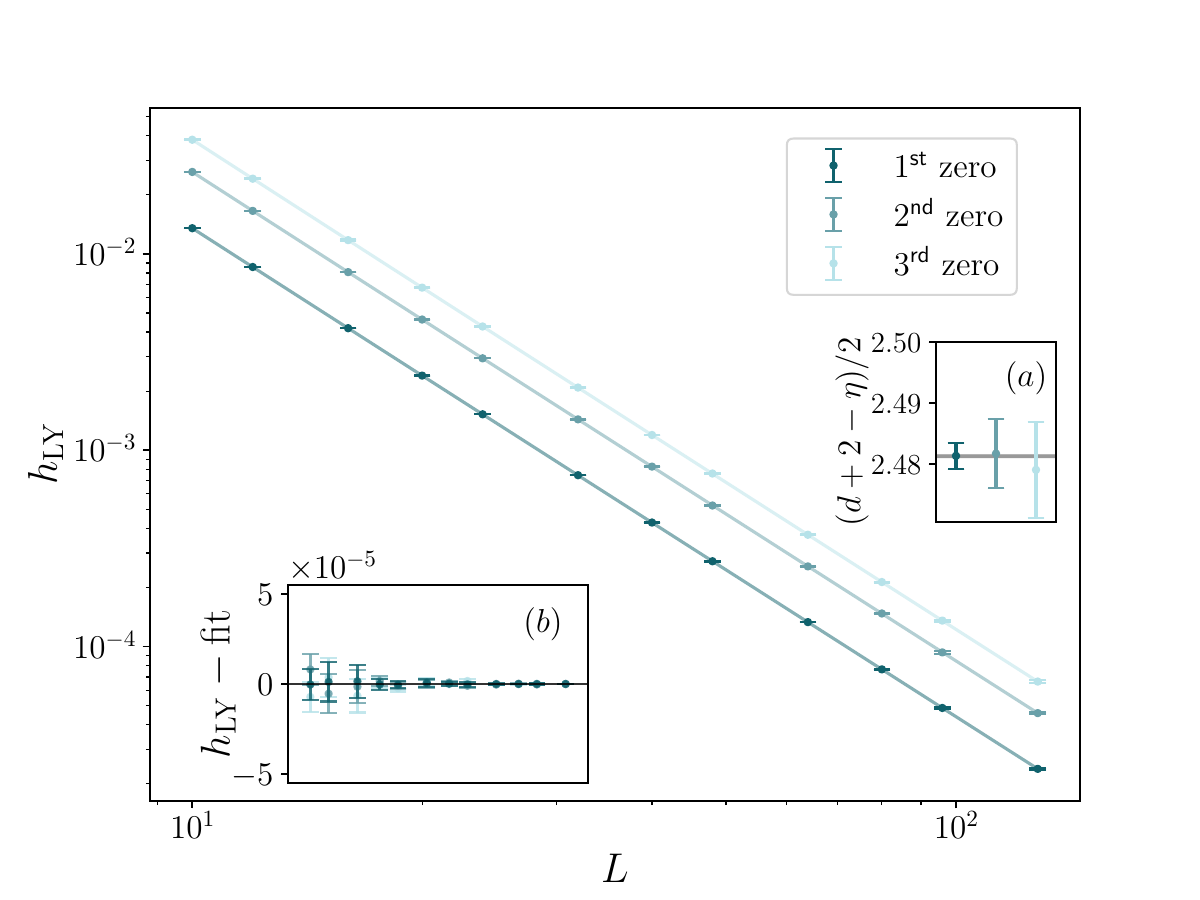}}
\caption{Main panel: The location on the imaginary axis of the first three Lee-Yang zeros of the Heisenberg bilayer in a complex N\'eel field at the critical point $J_{\perp}/J=2.5222$ and $\beta=L/2$. Panel (a): the exponent extracted from a fit of the data to the form $h_{\text{LY}}(L)\sim L^{-(d+2-\eta)/2}(1+ bL^{-\omega})$, where the fit is performed separately for each of the first three zeros.  The grey line is the best estimate of the exponent from the literature [\onlinecite{Campostrini2002:CritExpHeis}].  Panel (b): the raw data minus the fit.}
\label{fig:lybilayer}
\end{figure}

Now we investigate the critical scaling of the leading zeros of the Heisenberg bilayer as we approach the thermodynamic limit,  using finite-size scaling of the location of the zeros to extract critical exponents.  At a critical point, Lee-Yang zeros are expected to scale as $h_{\text{LY}}(L)\sim L^{-\beta\delta/\nu}$ [\onlinecite{Itzykson1983:DistributionZeros}], where, through scaling relations, the exponent can be expressed as $ \beta\delta / \nu = (d+2-\eta)/2$.  For the Heisenberg bilayer, we find it necessary to include corrections to scaling of the form $h_{\text{LY}}(L)\sim L^{-(d+2-\eta)/2}(1+ bL^{-\omega})$, as has been previously observed with Lee-Yang zeros for the 3D classical Heisenberg model [\onlinecite{Gordillo2013:LYZHeisenberg}].  Fig. (\ref{fig:lybilayer}) shows the first three zeros extracted from bilayer systems collected at $J=1, J_{\perp}=2.5222$ and $\beta=L/2$.  The extracted exponent is given in inset $(a)$, which agrees perfectly with the best estimate from classical Monte Carlo simulations [\onlinecite{Campostrini2002:CritExpHeis}].  Interestingly, we also find the correction to scaling $\omega \approx 1$ in agreement with the classical case [\onlinecite{Gordillo2013:LYZHeisenberg}].

{\em Deconfined criticality:}
Now that we have demonstrated the ability to extract critical exponents from the finite-size scaling of Lee-Yang zeros for the first time at a quantum critical point, we move on to a more exotic quantum spin model that shows criticality and emergent symmetry between two seemingly unrelated order parameters.

The $J$-$Q$ model [\onlinecite{Sandvik2007:JQ}] is given by the following Hamiltonian:
\begin{equation}
\label{eq:Hjq}
H_{JQ}= J\sum_{\langle i,j \rangle}(\vec{S}_{i} \cdot \vec{S}_{j} -\tfrac{1}{4}) - Q\sum_{\langle i,j,k,l \rangle}(\vec{S}_{i} \cdot \vec{S}_{j} -\tfrac{1}{4})(\vec{S}_{k} \cdot \vec{S}_{l} -\tfrac{1}{4}),
\end{equation}
where $J$ is the nearest-neighbor coupling on a square lattice and $Q$ is a four-spin interaction that acts on elementary plaquettes of the square lattice.  $Q$ can be thought of as a product of two $J$'s and the sum over $\langle i,j,k,l \rangle$ includes both $\hat{x}$ and $\hat{y}$ orientations to preserve lattice symmetries.
When $Q=0$ the $J$-$Q$ model reduces to the Heisenberg antiferromagnet, which exhibits long-range N\'eel order at $T=0$.  However, for large $Q$, magnetic order is destroyed by locally forming singlets that stack along columns of the square lattice, referred to as valence-bond solid (VBS) order, which breaks lattice translation symmetry.  At $J/Q \approx 0.045$ the system undergoes a seemingly continuous transition between these two distinct ordered phases, which belongs to a class of transitions known as deconfined quantum critical points (DQCP) [\onlinecite{Senthil2004:DQCPs}, \onlinecite{Senthil2004:BeyondLGW}].

The true nature of the phase transition in the $J$-$Q$ model has been the topic of continued debate.  While most studies observe a direct continuous transition between N\'eel and VBS order [\onlinecite{Sandvik2007:JQ,Lou2009:AFtoVBSsun,Sandvik2010:ContinuousJQlog,Harada2013:DQCPsmallN,Block2013:Fate,Shao2016:TwoLength,Sandvik2020:Consistent,Zhao2020:Helical}], other studies show evidence of a weak first-order transition [\onlinecite{Kuklov2008:DCPfirstorder,Jiang2008:FirstOrder,Chen2013:DCPflow,demidio2021diagnosing,Zhao2022:EEatDQC,song2023deconfined}].  The $J$-$Q$ model therefore provides us with an important case for the study of Lee-Yang zeros.

Having already developed the methodology to extract Lee-Yang zeros associated with fluctuations of the N\'eel order parameter, we would now like to develop an analogous treatment for the VBS order parameter.  Here we must introduce a complex-valued field that couples to the VBS order parameter, which is a field that preserves spin rotation symmetry but breaks lattice translation symmetry.  In order to do so, we make use of the $Q$ plaquette terms that are already present in the Hamiltonian, and we introduce the field $d\sum_{\langle i,j,k,l \rangle \in \hat{x}_e}P_{i,j,k,l} -d\sum_{\langle i,j,k,l \rangle \in \hat{x}_o}P_{i,j,k,l}$  where $P_{i,j,k,l} \equiv (\vec{S}_{i} \cdot \vec{S}_{j} -\tfrac{1}{4})(\vec{S}_{k} \cdot \vec{S}_{l} -\tfrac{1}{4})$ and $\hat{x}_e$ are all $\hat{x}$-oriented plaquettes on even columns of the square lattice and $\hat{x}_o$ are all $\hat{x}$-oriented plaquettes on odd columns.  $d$ is the coupling strength of the VBS field, which we will take to be complex-valued.

\begin{figure}[!t]
\centerline{\includegraphics[angle=0,width=0.95\columnwidth]{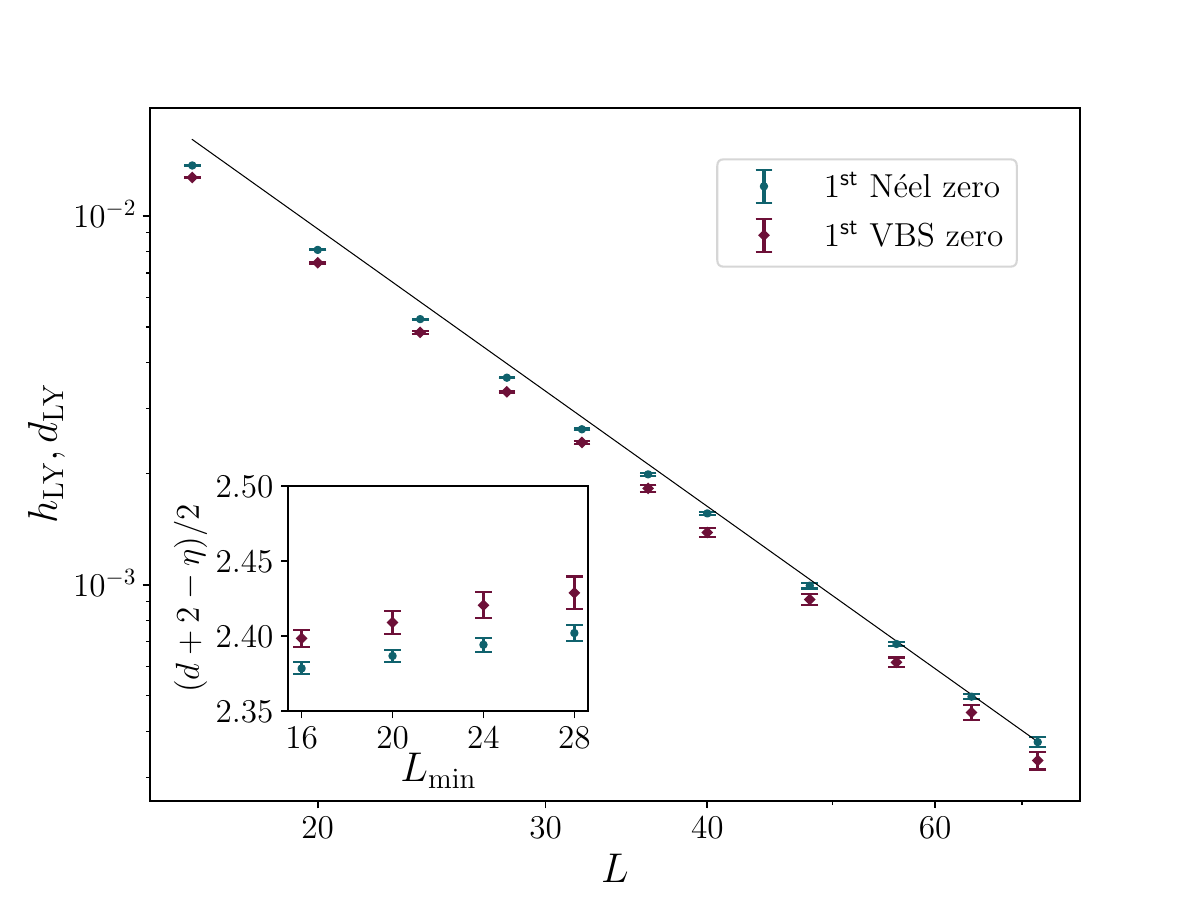}}
\caption{Main panel: the locations of the first Lee-Yang zeros in the $J$-$Q$ model at the critical point in two separate cases: with a complex N\'eel field and with a complex VBS field.  The thin black line shows the power law behavior expected at a first order transition with $\eta=0$. Inset:  The fitted exponent from a simple power law fit of the zeros to the form $\sim L^{-(d+2-\eta)/2}$ as a function of the smallest system size used in the fit.}
\label{fig:lyjq}
\end{figure}

Following the same prescription as the ratio formula with a complex N\'eel field, we have the ratio with a complex VBS field as follows:
\begin{equation}
\label{eq:zratd}
R_{0}(d)= \left\langle  \left(1-\frac{d}{Q}\right)^{n_{Q_{xe}}}  \left(1+\frac{d}{Q}\right)^{n_{Q_{xo}}} \right\rangle_{d=0}.
\end{equation}
Here $n_{Q_{xe}}$ is the number $\hat{x}$-oriented $Q$ matrix elements on even columns of the square lattice and $n_{Q_{xo}}$ is the same but for odd columns.  Again, we only require a histogram of the values $(n_{Q_{xe}},n_{Q_{xo}})$ obtained from QMC simulation of the pure $J$-$Q$ model without any fields, and Eqn. (\ref{eq:zratd}) allows for the statistically exact extraction of Lee-Yang zeros of the partition function with a complex VBS field.  For the $J$-$Q$ model we can simultaneously gather histograms for both the N\'eel zeros and VBS zeros in a single simulation.

In Fig. (\ref{fig:lyjq}) we plot the location of the Lee-Yang zeros $h_{\text{LY}},d_{\text{LY}}$ as a function of system size at the critical point $J=0.04502$, $Q=1$ with $\beta=L/2$.  Indeed, as with the N\'eel zeros, we find the nearest VBS zeros lie exactly on the imaginary axis: re$(h_{\text{LY}}) = $ re$(d_{\text{LY}}) = 0$.  For simplicity we plot only the location of the leading zeros.  In the $J$-$Q$ model, contrary to the Heisenberg bilayer, we achieve much better fits using only a simple power law with no corrections to scaling.  The extracted exponents are far below the values of the bilayer, signaling a much larger value of $\eta$ here, with our values in the range of previous studies.  We do, however, observe drifting of the critical exponents, as shown in the inset, which is also similar to what was observed in other studies [\onlinecite{Harada2013:DQCPsmallN}, \onlinecite{Nahum2015:DQCloop}].

{\em Combined N\'eel and VBS fields:}
The transition in the $J$-$Q$ model is frequently described in terms of an emergent SO(5) symmetry that arises between the N\'eel and VBS order parameters exactly at the critical point.  From this point of view, it is interesting to ask the how the zeros of the partition function are distributed in the presence of combined N\'eel and VBS fields.  Fortunately, our formulation allows us to simply compute the ratio when both fields are nonzero, again, while the actual simulation is carried out in zero field.  We can straightforwardly define the combined ratio as:
\begin{equation}
\label{eq:Rlyzhd}
\begin{split}
R_{0}(h,d) &= \left\langle \left(1-\frac{2 h}{N_c J}\right)^{n_{\uparrow\downarrow}}  \left(1+\frac{2 h}{N_c J}\right)^{n_{\downarrow\uparrow}}\right. \\
&\left. \quad \quad  \left(1-\frac{d}{Q}\right)^{n_{Q_{xe}}}  \left(1+\frac{d}{Q}\right)^{n_{Q_{xo}}} \right\rangle_{h=0,d=0}.
\end{split}
\end{equation}

\begin{figure}[t]
\centerline{\includegraphics[angle=0,width=0.95\columnwidth]{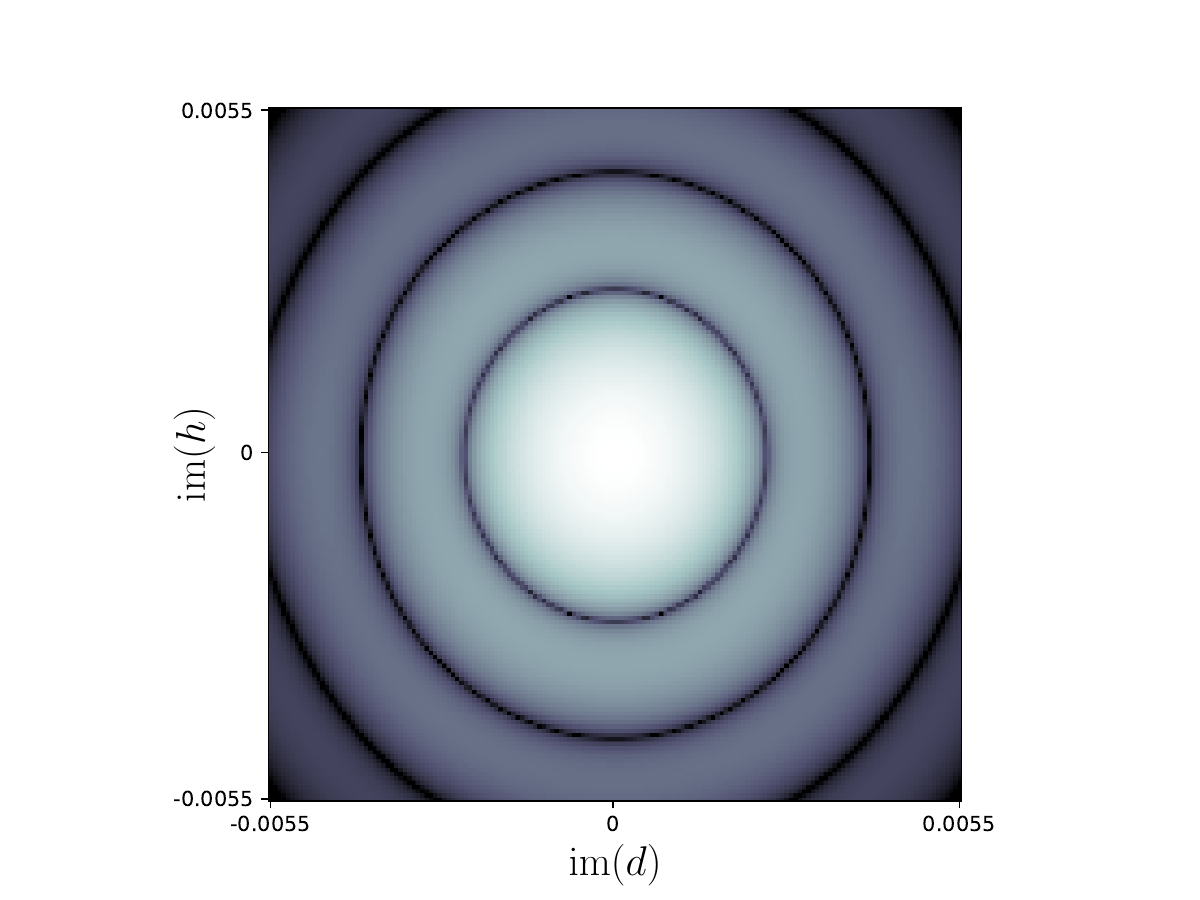}}
\caption{A density plot of $-\ln(|R_0(h,d)|)$ for the $J$-$Q$ model at the critical coupling with $L=32$.  The dark lines indicate zeros of the partition function in the $J$-$Q$ model subjected to combined N\'eel and VBS fields.}
\label{fig:jqmap}
\end{figure}

In Fig. (\ref{fig:jqmap}) we plot $-\ln(|R_0(h,d)|)$ for purely imaginary values of $h$ and $d$ at the critical point.  Interestingly, we see that the partition function zeros form extended rings in the plane (im($h$),im($d$)).  On either side of the transition the ovals become elongated in either the vertical or horizontal direction (see supplemental materials).  Our conclusion is that at the critical point these rings shrink down to the origin in the thermodynamic limit, whereas on either side of the transition they become squashed in either the horizontal or vertical direction.  It is interesting to note that although the values of $J$ and $Q$ differ by an order of magnitude at the transition, the lines of zeros are nearly circular in terms of $h$ and $d$.

{\em Conclusions:}
We have presented the first large-scale QMC computations of Lee-Yang zeros in quantum spin systems.  Our formulation is simple to implement, in that it only requires gathering a histogram of the number of certain types of matrix elements that appear during standard QMC simulations in the absence of any external fields.  Furthermore, we find that the location of Lee-Yang zeros computed in this way is extremely precise, easily allowing for the extraction of the the first three zeros for the critical Heisenberg bilayer on system sizes up to $L=128$ and excellent agreement with known critical exponents.  We also have demonstrated the generality of this approach by using it to extract Lee-Yang zeros associated with critical N\'eel, VBS and combined N\'eel-VBS fluctuations in the $J$-$Q$ model, the emblematic model of deconfined criticality.

Moving forward, we envision that this technique could be applied to extract properties of non-unitary critical points that occur at finite imaginary field values.  Specifically, with the partition function ratio in hand, it allows one to compute expectation values in the ensemble with a complex field.  This seems promising given that the ratio is computed to high precision in this framework.  On a speculative note, these types of studies could aid in understanding lattice models with pseudo-critical behavior, where the true critical point lies in the complex plane but with a small imaginary component, as has been proposed to explain anomalous behavior in models of deconfined criticality [\onlinecite{Nahum2015:DQCloop,Gorbenko2018:Walking,Ma2020:Pseudo,Nahum2020:Quasi,song2023deconfined}].

{\em Acknowledgements:} We thank Román Orús for his support for this project.  Computing resources were used from the XSEDE allocation NSF DMR-130040 using the Expanse cluster at the San Diego Supercomputer Center.

\bibliography{LeeYangQMC}

\begin{thebibliography}{55}%
\makeatletter
\providecommand \@ifxundefined [1]{%
 \@ifx{#1\undefined}
}%
\providecommand \@ifnum [1]{%
 \ifnum #1\expandafter \@firstoftwo
 \else \expandafter \@secondoftwo
 \fi
}%
\providecommand \@ifx [1]{%
 \ifx #1\expandafter \@firstoftwo
 \else \expandafter \@secondoftwo
 \fi
}%
\providecommand \natexlab [1]{#1}%
\providecommand \enquote  [1]{``#1''}%
\providecommand \bibnamefont  [1]{#1}%
\providecommand \bibfnamefont [1]{#1}%
\providecommand \citenamefont [1]{#1}%
\providecommand \href@noop [0]{\@secondoftwo}%
\providecommand \href [0]{\begingroup \@sanitize@url \@href}%
\providecommand \@href[1]{\@@startlink{#1}\@@href}%
\providecommand \@@href[1]{\endgroup#1\@@endlink}%
\providecommand \@sanitize@url [0]{\catcode `\\12\catcode `\$12\catcode
  `\&12\catcode `\#12\catcode `\^12\catcode `\_12\catcode `\%12\relax}%
\providecommand \@@startlink[1]{}%
\providecommand \@@endlink[0]{}%
\providecommand \url  [0]{\begingroup\@sanitize@url \@url }%
\providecommand \@url [1]{\endgroup\@href {#1}{\urlprefix }}%
\providecommand \urlprefix  [0]{URL }%
\providecommand \Eprint [0]{\href }%
\providecommand \doibase [0]{http://dx.doi.org/}%
\providecommand \selectlanguage [0]{\@gobble}%
\providecommand \bibinfo  [0]{\@secondoftwo}%
\providecommand \bibfield  [0]{\@secondoftwo}%
\providecommand \translation [1]{[#1]}%
\providecommand \BibitemOpen [0]{}%
\providecommand \bibitemStop [0]{}%
\providecommand \bibitemNoStop [0]{.\EOS\space}%
\providecommand \EOS [0]{\spacefactor3000\relax}%
\providecommand \BibitemShut  [1]{\csname bibitem#1\endcsname}%
\let\auto@bib@innerbib\@empty
%</preamble>
\bibitem [{\citenamefont {Yang}\ and\ \citenamefont
  {Lee}(1952)}]{YangLee1952:StatTheoryI}%
  \BibitemOpen
  \bibfield  {author} {\bibinfo {author} {\bibfnamefont {C.~N.}\ \bibnamefont
  {Yang}}\ and\ \bibinfo {author} {\bibfnamefont {T.~D.}\ \bibnamefont {Lee}},\
  }\bibfield  {title} {\enquote {\bibinfo {title} {Statistical theory of
  equations of state and phase transitions. i. theory of condensation},}\
  }\href {\doibase 10.1103/PhysRev.87.404} {\bibfield  {journal} {\bibinfo
  {journal} {Phys. Rev.}\ }\textbf {\bibinfo {volume} {87}},\ \bibinfo {pages}
  {404--409} (\bibinfo {year} {1952})}\BibitemShut {NoStop}%
\bibitem [{\citenamefont {Lee}\ and\ \citenamefont
  {Yang}(1952)}]{LeeYang1952:StatTheoryII}%
  \BibitemOpen
  \bibfield  {author} {\bibinfo {author} {\bibfnamefont {T.~D.}\ \bibnamefont
  {Lee}}\ and\ \bibinfo {author} {\bibfnamefont {C.~N.}\ \bibnamefont {Yang}},\
  }\bibfield  {title} {\enquote {\bibinfo {title} {Statistical theory of
  equations of state and phase transitions. ii. lattice gas and ising model},}\
  }\href {\doibase 10.1103/PhysRev.87.410} {\bibfield  {journal} {\bibinfo
  {journal} {Phys. Rev.}\ }\textbf {\bibinfo {volume} {87}},\ \bibinfo {pages}
  {410--419} (\bibinfo {year} {1952})}\BibitemShut {NoStop}%
\bibitem [{\citenamefont {Fisher}(1978)}]{Fisher1978:LYEphi3}%
  \BibitemOpen
  \bibfield  {author} {\bibinfo {author} {\bibfnamefont {Michael~E.}\
  \bibnamefont {Fisher}},\ }\bibfield  {title} {\enquote {\bibinfo {title}
  {Yang-lee edge singularity and ${\ensuremath{\phi}}^{3}$ field theory},}\
  }\href {\doibase 10.1103/PhysRevLett.40.1610} {\bibfield  {journal} {\bibinfo
   {journal} {Phys. Rev. Lett.}\ }\textbf {\bibinfo {volume} {40}},\ \bibinfo
  {pages} {1610--1613} (\bibinfo {year} {1978})}\BibitemShut {NoStop}%
\bibitem [{\citenamefont {Cardy}(1985)}]{Cardy1985:ConformalLYE}%
  \BibitemOpen
  \bibfield  {author} {\bibinfo {author} {\bibfnamefont {John~L.}\ \bibnamefont
  {Cardy}},\ }\bibfield  {title} {\enquote {\bibinfo {title} {Conformal
  invariance and the yang-lee edge singularity in two dimensions},}\ }\href
  {\doibase 10.1103/PhysRevLett.54.1354} {\bibfield  {journal} {\bibinfo
  {journal} {Phys. Rev. Lett.}\ }\textbf {\bibinfo {volume} {54}},\ \bibinfo
  {pages} {1354--1356} (\bibinfo {year} {1985})}\BibitemShut {NoStop}%
\bibitem [{\citenamefont {Binek}(1998)}]{Binek1998:LYZmag}%
  \BibitemOpen
  \bibfield  {author} {\bibinfo {author} {\bibfnamefont {Ch.}\ \bibnamefont
  {Binek}},\ }\bibfield  {title} {\enquote {\bibinfo {title} {Density of zeros
  on the lee-yang circle obtained from magnetization data of a two-dimensional
  ising ferromagnet},}\ }\href {\doibase 10.1103/PhysRevLett.81.5644}
  {\bibfield  {journal} {\bibinfo  {journal} {Phys. Rev. Lett.}\ }\textbf
  {\bibinfo {volume} {81}},\ \bibinfo {pages} {5644--5647} (\bibinfo {year}
  {1998})}\BibitemShut {NoStop}%
\bibitem [{\citenamefont {Peng}\ \emph {et~al.}(2015)\citenamefont {Peng},
  \citenamefont {Zhou}, \citenamefont {Wei}, \citenamefont {Cui}, \citenamefont
  {Du},\ and\ \citenamefont {Liu}}]{Peng2015:ExpLYZ}%
  \BibitemOpen
  \bibfield  {author} {\bibinfo {author} {\bibfnamefont {Xinhua}\ \bibnamefont
  {Peng}}, \bibinfo {author} {\bibfnamefont {Hui}\ \bibnamefont {Zhou}},
  \bibinfo {author} {\bibfnamefont {Bo-Bo}\ \bibnamefont {Wei}}, \bibinfo
  {author} {\bibfnamefont {Jiangyu}\ \bibnamefont {Cui}}, \bibinfo {author}
  {\bibfnamefont {Jiangfeng}\ \bibnamefont {Du}}, \ and\ \bibinfo {author}
  {\bibfnamefont {Ren-Bao}\ \bibnamefont {Liu}},\ }\bibfield  {title} {\enquote
  {\bibinfo {title} {Experimental observation of lee-yang zeros},}\ }\href
  {\doibase 10.1103/PhysRevLett.114.010601} {\bibfield  {journal} {\bibinfo
  {journal} {Phys. Rev. Lett.}\ }\textbf {\bibinfo {volume} {114}},\ \bibinfo
  {pages} {010601} (\bibinfo {year} {2015})}\BibitemShut {NoStop}%
\bibitem [{\citenamefont {Brandner}\ \emph {et~al.}(2017)\citenamefont
  {Brandner}, \citenamefont {Maisi}, \citenamefont {Pekola}, \citenamefont
  {Garrahan},\ and\ \citenamefont {Flindt}}]{Brandner2017:ExpDynLYZ}%
  \BibitemOpen
  \bibfield  {author} {\bibinfo {author} {\bibfnamefont {Kay}\ \bibnamefont
  {Brandner}}, \bibinfo {author} {\bibfnamefont {Ville~F.}\ \bibnamefont
  {Maisi}}, \bibinfo {author} {\bibfnamefont {Jukka~P.}\ \bibnamefont
  {Pekola}}, \bibinfo {author} {\bibfnamefont {Juan~P.}\ \bibnamefont
  {Garrahan}}, \ and\ \bibinfo {author} {\bibfnamefont {Christian}\
  \bibnamefont {Flindt}},\ }\bibfield  {title} {\enquote {\bibinfo {title}
  {Experimental determination of dynamical lee-yang zeros},}\ }\href {\doibase
  10.1103/PhysRevLett.118.180601} {\bibfield  {journal} {\bibinfo  {journal}
  {Phys. Rev. Lett.}\ }\textbf {\bibinfo {volume} {118}},\ \bibinfo {pages}
  {180601} (\bibinfo {year} {2017})}\BibitemShut {NoStop}%
\bibitem [{\citenamefont {Francis}\ \emph {et~al.}(2021)\citenamefont
  {Francis}, \citenamefont {Zhu}, \citenamefont {Alderete}, \citenamefont
  {Johri}, \citenamefont {Xiao}, \citenamefont {Freericks}, \citenamefont
  {Monroe}, \citenamefont {Linke},\ and\ \citenamefont
  {Kemper}}]{Akhil2021:QCLYZ}%
  \BibitemOpen
  \bibfield  {author} {\bibinfo {author} {\bibfnamefont {Akhil}\ \bibnamefont
  {Francis}}, \bibinfo {author} {\bibfnamefont {Daiwei}\ \bibnamefont {Zhu}},
  \bibinfo {author} {\bibfnamefont {Cinthia~Huerta}\ \bibnamefont {Alderete}},
  \bibinfo {author} {\bibfnamefont {Sonika}\ \bibnamefont {Johri}}, \bibinfo
  {author} {\bibfnamefont {Xiao}\ \bibnamefont {Xiao}}, \bibinfo {author}
  {\bibfnamefont {James~K.}\ \bibnamefont {Freericks}}, \bibinfo {author}
  {\bibfnamefont {Christopher}\ \bibnamefont {Monroe}}, \bibinfo {author}
  {\bibfnamefont {Norbert~M.}\ \bibnamefont {Linke}}, \ and\ \bibinfo {author}
  {\bibfnamefont {Alexander~F.}\ \bibnamefont {Kemper}},\ }\bibfield  {title}
  {\enquote {\bibinfo {title} {Many-body thermodynamics on quantum computers
  via partition function zeros},}\ }\href {\doibase 10.1126/sciadv.abf2447}
  {\bibfield  {journal} {\bibinfo  {journal} {Science Advances}\ }\textbf
  {\bibinfo {volume} {7}},\ \bibinfo {pages} {eabf2447} (\bibinfo {year}
  {2021})},\ \Eprint
  {http://arxiv.org/abs/https://www.science.org/doi/pdf/10.1126/sciadv.abf2447}
  {https://www.science.org/doi/pdf/10.1126/sciadv.abf2447} \BibitemShut
  {NoStop}%
\bibitem [{\citenamefont {Lee}(2013)}]{Lee2013:LYZprotein}%
  \BibitemOpen
  \bibfield  {author} {\bibinfo {author} {\bibfnamefont {Julian}\ \bibnamefont
  {Lee}},\ }\bibfield  {title} {\enquote {\bibinfo {title} {Exact partition
  function zeros of the wako-sait\^o-mu\~noz-eaton protein model},}\ }\href
  {\doibase 10.1103/PhysRevLett.110.248101} {\bibfield  {journal} {\bibinfo
  {journal} {Phys. Rev. Lett.}\ }\textbf {\bibinfo {volume} {110}},\ \bibinfo
  {pages} {248101} (\bibinfo {year} {2013})}\BibitemShut {NoStop}%
\bibitem [{\citenamefont {Deger}\ \emph {et~al.}(2018)\citenamefont {Deger},
  \citenamefont {Brandner},\ and\ \citenamefont
  {Flindt}}]{Deger2018:LYZzipper}%
  \BibitemOpen
  \bibfield  {author} {\bibinfo {author} {\bibfnamefont {Aydin}\ \bibnamefont
  {Deger}}, \bibinfo {author} {\bibfnamefont {Kay}\ \bibnamefont {Brandner}}, \
  and\ \bibinfo {author} {\bibfnamefont {Christian}\ \bibnamefont {Flindt}},\
  }\bibfield  {title} {\enquote {\bibinfo {title} {Lee-yang zeros and
  large-deviation statistics of a molecular zipper},}\ }\href {\doibase
  10.1103/PhysRevE.97.012115} {\bibfield  {journal} {\bibinfo  {journal} {Phys.
  Rev. E}\ }\textbf {\bibinfo {volume} {97}},\ \bibinfo {pages} {012115}
  (\bibinfo {year} {2018})}\BibitemShut {NoStop}%
\bibitem [{\citenamefont {Majumdar}\ and\ \citenamefont
  {Bhattacharjee}(2020)}]{Majumdar2020:LYZcumulant}%
  \BibitemOpen
  \bibfield  {author} {\bibinfo {author} {\bibfnamefont {Debjyoti}\
  \bibnamefont {Majumdar}}\ and\ \bibinfo {author} {\bibfnamefont
  {Somendra~M.}\ \bibnamefont {Bhattacharjee}},\ }\bibfield  {title} {\enquote
  {\bibinfo {title} {Zeros of partition function for continuous phase
  transitions using cumulants},}\ }\href {\doibase
  https://doi.org/10.1016/j.physa.2020.124263} {\bibfield  {journal} {\bibinfo
  {journal} {Physica A: Statistical Mechanics and its Applications}\ }\textbf
  {\bibinfo {volume} {553}},\ \bibinfo {pages} {124263} (\bibinfo {year}
  {2020})}\BibitemShut {NoStop}%
\bibitem [{\citenamefont {Nagata}\ \emph {et~al.}(2015)\citenamefont {Nagata},
  \citenamefont {Kashiwa}, \citenamefont {Nakamura},\ and\ \citenamefont
  {Nishigaki}}]{Nagata2015:LYZQCD}%
  \BibitemOpen
  \bibfield  {author} {\bibinfo {author} {\bibfnamefont {Keitaro}\ \bibnamefont
  {Nagata}}, \bibinfo {author} {\bibfnamefont {Kouji}\ \bibnamefont {Kashiwa}},
  \bibinfo {author} {\bibfnamefont {Atsushi}\ \bibnamefont {Nakamura}}, \ and\
  \bibinfo {author} {\bibfnamefont {Shinsuke~M.}\ \bibnamefont {Nishigaki}},\
  }\bibfield  {title} {\enquote {\bibinfo {title} {Lee-yang zero distribution
  of high temperature qcd and the roberge-weiss phase transition},}\ }\href
  {\doibase 10.1103/PhysRevD.91.094507} {\bibfield  {journal} {\bibinfo
  {journal} {Phys. Rev. D}\ }\textbf {\bibinfo {volume} {91}},\ \bibinfo
  {pages} {094507} (\bibinfo {year} {2015})}\BibitemShut {NoStop}%
\bibitem [{\citenamefont {Bena}\ \emph {et~al.}(2005)\citenamefont {Bena},
  \citenamefont {Droz},\ and\ \citenamefont
  {Lipowski}}]{Bena2005:YangLeeReview}%
  \BibitemOpen
  \bibfield  {author} {\bibinfo {author} {\bibfnamefont {Ioana}\ \bibnamefont
  {Bena}}, \bibinfo {author} {\bibfnamefont {Michel}\ \bibnamefont {Droz}}, \
  and\ \bibinfo {author} {\bibfnamefont {Adam}\ \bibnamefont {Lipowski}},\
  }\bibfield  {title} {\enquote {\bibinfo {title} {Statistical mechanics of
  equilibrium and nonequilibrium phase transitions: The yang--lee formalism},}\
  }\href {\doibase 10.1142/S0217979205032759} {\bibfield  {journal} {\bibinfo
  {journal} {International Journal of Modern Physics B}\ }\textbf {\bibinfo
  {volume} {19}},\ \bibinfo {pages} {4269--4329} (\bibinfo {year} {2005})},\
  \Eprint {http://arxiv.org/abs/https://doi.org/10.1142/S0217979205032759}
  {https://doi.org/10.1142/S0217979205032759} \BibitemShut {NoStop}%
\bibitem [{\citenamefont {Ferrenberg}\ and\ \citenamefont
  {Swendsen}(1988)}]{Ferrenberg1988:MCtransition}%
  \BibitemOpen
  \bibfield  {author} {\bibinfo {author} {\bibfnamefont {Alan~M.}\ \bibnamefont
  {Ferrenberg}}\ and\ \bibinfo {author} {\bibfnamefont {Robert~H.}\
  \bibnamefont {Swendsen}},\ }\bibfield  {title} {\enquote {\bibinfo {title}
  {New monte carlo technique for studying phase transitions},}\ }\href
  {\doibase 10.1103/PhysRevLett.61.2635} {\bibfield  {journal} {\bibinfo
  {journal} {Phys. Rev. Lett.}\ }\textbf {\bibinfo {volume} {61}},\ \bibinfo
  {pages} {2635--2638} (\bibinfo {year} {1988})}\BibitemShut {NoStop}%
\bibitem [{\citenamefont {Kenna}\ and\ \citenamefont
  {Lang}(1993)}]{Kenna1993:LYZlogphi4}%
  \BibitemOpen
  \bibfield  {author} {\bibinfo {author} {\bibfnamefont {R.}~\bibnamefont
  {Kenna}}\ and\ \bibinfo {author} {\bibfnamefont {C.B.}\ \bibnamefont
  {Lang}},\ }\bibfield  {title} {\enquote {\bibinfo {title} {Lee-yang zeroes
  and logarithmic corrections in the {$\Phi^{4}_4$} theory},}\ }\href {\doibase
  https://doi.org/10.1016/0920-5632(93)90305-P} {\bibfield  {journal} {\bibinfo
   {journal} {Nuclear Physics B - Proceedings Supplements}\ }\textbf {\bibinfo
  {volume} {30}},\ \bibinfo {pages} {697--700} (\bibinfo {year} {1993})},\
  \bibinfo {note} {proceedings of the International Symposium on}\BibitemShut
  {NoStop}%
\bibitem [{\citenamefont {Kenna}\ and\ \citenamefont
  {Lang}(1994)}]{Kenna1994:LYZ4dIsing}%
  \BibitemOpen
  \bibfield  {author} {\bibinfo {author} {\bibfnamefont {R.}~\bibnamefont
  {Kenna}}\ and\ \bibinfo {author} {\bibfnamefont {C.~B.}\ \bibnamefont
  {Lang}},\ }\bibfield  {title} {\enquote {\bibinfo {title} {Scaling and
  density of lee-yang zeros in the four-dimensional ising model},}\ }\href
  {\doibase 10.1103/PhysRevE.49.5012} {\bibfield  {journal} {\bibinfo
  {journal} {Phys. Rev. E}\ }\textbf {\bibinfo {volume} {49}},\ \bibinfo
  {pages} {5012--5017} (\bibinfo {year} {1994})}\BibitemShut {NoStop}%
\bibitem [{\citenamefont {Flindt}\ and\ \citenamefont
  {Garrahan}(2013)}]{Flindt2013:TrajPhaseTransLYZ}%
  \BibitemOpen
  \bibfield  {author} {\bibinfo {author} {\bibfnamefont {Christian}\
  \bibnamefont {Flindt}}\ and\ \bibinfo {author} {\bibfnamefont {Juan~P.}\
  \bibnamefont {Garrahan}},\ }\bibfield  {title} {\enquote {\bibinfo {title}
  {Trajectory phase transitions, lee-yang zeros, and high-order cumulants in
  full counting statistics},}\ }\href {\doibase 10.1103/PhysRevLett.110.050601}
  {\bibfield  {journal} {\bibinfo  {journal} {Phys. Rev. Lett.}\ }\textbf
  {\bibinfo {volume} {110}},\ \bibinfo {pages} {050601} (\bibinfo {year}
  {2013})}\BibitemShut {NoStop}%
\bibitem [{\citenamefont {Deger}\ \emph {et~al.}(2020)\citenamefont {Deger},
  \citenamefont {Brange},\ and\ \citenamefont
  {Flindt}}]{Deger2020:LYZcumulantIsing}%
  \BibitemOpen
  \bibfield  {author} {\bibinfo {author} {\bibfnamefont {Aydin}\ \bibnamefont
  {Deger}}, \bibinfo {author} {\bibfnamefont {Fredrik}\ \bibnamefont {Brange}},
  \ and\ \bibinfo {author} {\bibfnamefont {Christian}\ \bibnamefont {Flindt}},\
  }\bibfield  {title} {\enquote {\bibinfo {title} {Lee-yang theory, high
  cumulants, and large-deviation statistics of the magnetization in the ising
  model},}\ }\href {\doibase 10.1103/PhysRevB.102.174418} {\bibfield  {journal}
  {\bibinfo  {journal} {Phys. Rev. B}\ }\textbf {\bibinfo {volume} {102}},\
  \bibinfo {pages} {174418} (\bibinfo {year} {2020})}\BibitemShut {NoStop}%
\bibitem [{\citenamefont {Shimizu}\ and\ \citenamefont
  {Kuramashi}(2014{\natexlab{a}})}]{Shimizu2014:GrassmanTNSchwinger}%
  \BibitemOpen
  \bibfield  {author} {\bibinfo {author} {\bibfnamefont {Yuya}\ \bibnamefont
  {Shimizu}}\ and\ \bibinfo {author} {\bibfnamefont {Yoshinobu}\ \bibnamefont
  {Kuramashi}},\ }\bibfield  {title} {\enquote {\bibinfo {title} {Grassmann
  tensor renormalization group approach to one-flavor lattice schwinger
  model},}\ }\href {\doibase 10.1103/PhysRevD.90.014508} {\bibfield  {journal}
  {\bibinfo  {journal} {Phys. Rev. D}\ }\textbf {\bibinfo {volume} {90}},\
  \bibinfo {pages} {014508} (\bibinfo {year} {2014}{\natexlab{a}})}\BibitemShut
  {NoStop}%
\bibitem [{\citenamefont {Shimizu}\ and\ \citenamefont
  {Kuramashi}(2014{\natexlab{b}})}]{Shimizu2014:GrassmanTNCriticalSchwinger}%
  \BibitemOpen
  \bibfield  {author} {\bibinfo {author} {\bibfnamefont {Yuya}\ \bibnamefont
  {Shimizu}}\ and\ \bibinfo {author} {\bibfnamefont {Yoshinobu}\ \bibnamefont
  {Kuramashi}},\ }\bibfield  {title} {\enquote {\bibinfo {title} {Critical
  behavior of the lattice schwinger model with a topological term at
  $\ensuremath{\theta}=\ensuremath{\pi}$ using the grassmann tensor
  renormalization group},}\ }\href {\doibase 10.1103/PhysRevD.90.074503}
  {\bibfield  {journal} {\bibinfo  {journal} {Phys. Rev. D}\ }\textbf {\bibinfo
  {volume} {90}},\ \bibinfo {pages} {074503} (\bibinfo {year}
  {2014}{\natexlab{b}})}\BibitemShut {NoStop}%
\bibitem [{\citenamefont {Garc\'{\i}a-Saez}\ and\ \citenamefont
  {Wei}(2015)}]{GarciaSaez2015:LYZTN}%
  \BibitemOpen
  \bibfield  {author} {\bibinfo {author} {\bibfnamefont {Artur}\ \bibnamefont
  {Garc\'{\i}a-Saez}}\ and\ \bibinfo {author} {\bibfnamefont {Tzu-Chieh}\
  \bibnamefont {Wei}},\ }\bibfield  {title} {\enquote {\bibinfo {title}
  {Density of yang-lee zeros in the thermodynamic limit from tensor network
  methods},}\ }\href {\doibase 10.1103/PhysRevB.92.125132} {\bibfield
  {journal} {\bibinfo  {journal} {Phys. Rev. B}\ }\textbf {\bibinfo {volume}
  {92}},\ \bibinfo {pages} {125132} (\bibinfo {year} {2015})}\BibitemShut
  {NoStop}%
\bibitem [{\citenamefont {Hong}\ and\ \citenamefont
  {Kim}(2022)}]{Hong2022:TNLYZXY}%
  \BibitemOpen
  \bibfield  {author} {\bibinfo {author} {\bibfnamefont {Seongpyo}\
  \bibnamefont {Hong}}\ and\ \bibinfo {author} {\bibfnamefont {Dong-Hee}\
  \bibnamefont {Kim}},\ }\bibfield  {title} {\enquote {\bibinfo {title} {Tensor
  network calculation of the logarithmic correction exponent in the xy
  model},}\ }\href {\doibase 10.7566/JPSJ.91.084003} {\bibfield  {journal}
  {\bibinfo  {journal} {Journal of the Physical Society of Japan}\ }\textbf
  {\bibinfo {volume} {91}},\ \bibinfo {pages} {084003} (\bibinfo {year}
  {2022})},\ \Eprint
  {http://arxiv.org/abs/https://doi.org/10.7566/JPSJ.91.084003}
  {https://doi.org/10.7566/JPSJ.91.084003} \BibitemShut {NoStop}%
\bibitem [{\citenamefont {Heyl}\ \emph {et~al.}(2013)\citenamefont {Heyl},
  \citenamefont {Polkovnikov},\ and\ \citenamefont
  {Kehrein}}]{Heyl2013:DynamicalQPT}%
  \BibitemOpen
  \bibfield  {author} {\bibinfo {author} {\bibfnamefont {M.}~\bibnamefont
  {Heyl}}, \bibinfo {author} {\bibfnamefont {A.}~\bibnamefont {Polkovnikov}}, \
  and\ \bibinfo {author} {\bibfnamefont {S.}~\bibnamefont {Kehrein}},\
  }\bibfield  {title} {\enquote {\bibinfo {title} {Dynamical quantum phase
  transitions in the transverse-field ising model},}\ }\href {\doibase
  10.1103/PhysRevLett.110.135704} {\bibfield  {journal} {\bibinfo  {journal}
  {Phys. Rev. Lett.}\ }\textbf {\bibinfo {volume} {110}},\ \bibinfo {pages}
  {135704} (\bibinfo {year} {2013})}\BibitemShut {NoStop}%
\bibitem [{\citenamefont {Ananikian}\ \emph {et~al.}(2014)\citenamefont
  {Ananikian}, \citenamefont {Hovhannisyan},\ and\ \citenamefont
  {Kenna}}]{Ananikian2014:Diamond}%
  \BibitemOpen
  \bibfield  {author} {\bibinfo {author} {\bibfnamefont {N.S.}\ \bibnamefont
  {Ananikian}}, \bibinfo {author} {\bibfnamefont {V.V.}\ \bibnamefont
  {Hovhannisyan}}, \ and\ \bibinfo {author} {\bibfnamefont {R.}~\bibnamefont
  {Kenna}},\ }\bibfield  {title} {\enquote {\bibinfo {title} {Partition
  function zeros of the antiferromagnetic spin-12 ising--heisenberg model on a
  diamond chain},}\ }\href {\doibase
  https://doi.org/10.1016/j.physa.2013.11.017} {\bibfield  {journal} {\bibinfo
  {journal} {Physica A: Statistical Mechanics and its Applications}\ }\textbf
  {\bibinfo {volume} {396}},\ \bibinfo {pages} {51--60} (\bibinfo {year}
  {2014})}\BibitemShut {NoStop}%
\bibitem [{\citenamefont {Peotta}\ \emph {et~al.}(2021)\citenamefont {Peotta},
  \citenamefont {Brange}, \citenamefont {Deger}, \citenamefont {Ojanen},\ and\
  \citenamefont {Flindt}}]{Peotta2021:DynamicalQPT}%
  \BibitemOpen
  \bibfield  {author} {\bibinfo {author} {\bibfnamefont {Sebastiano}\
  \bibnamefont {Peotta}}, \bibinfo {author} {\bibfnamefont {Fredrik}\
  \bibnamefont {Brange}}, \bibinfo {author} {\bibfnamefont {Aydin}\
  \bibnamefont {Deger}}, \bibinfo {author} {\bibfnamefont {Teemu}\ \bibnamefont
  {Ojanen}}, \ and\ \bibinfo {author} {\bibfnamefont {Christian}\ \bibnamefont
  {Flindt}},\ }\bibfield  {title} {\enquote {\bibinfo {title} {Determination of
  dynamical quantum phase transitions in strongly correlated many-body systems
  using loschmidt cumulants},}\ }\href {\doibase 10.1103/PhysRevX.11.041018}
  {\bibfield  {journal} {\bibinfo  {journal} {Phys. Rev. X}\ }\textbf {\bibinfo
  {volume} {11}},\ \bibinfo {pages} {041018} (\bibinfo {year}
  {2021})}\BibitemShut {NoStop}%
\bibitem [{\citenamefont {Kist}\ \emph {et~al.}(2021)\citenamefont {Kist},
  \citenamefont {Lado},\ and\ \citenamefont {Flindt}}]{Kist2021:LYTheoryQMB}%
  \BibitemOpen
  \bibfield  {author} {\bibinfo {author} {\bibfnamefont {Timo}\ \bibnamefont
  {Kist}}, \bibinfo {author} {\bibfnamefont {Jose~L.}\ \bibnamefont {Lado}}, \
  and\ \bibinfo {author} {\bibfnamefont {Christian}\ \bibnamefont {Flindt}},\
  }\bibfield  {title} {\enquote {\bibinfo {title} {Lee-yang theory of
  criticality in interacting quantum many-body systems},}\ }\href {\doibase
  10.1103/PhysRevResearch.3.033206} {\bibfield  {journal} {\bibinfo  {journal}
  {Phys. Rev. Res.}\ }\textbf {\bibinfo {volume} {3}},\ \bibinfo {pages}
  {033206} (\bibinfo {year} {2021})}\BibitemShut {NoStop}%
\bibitem [{\citenamefont {Vecsei}\ \emph {et~al.}(2022)\citenamefont {Vecsei},
  \citenamefont {Lado},\ and\ \citenamefont
  {Flindt}}]{Vecsei2022:LYZ2DquantumIsing}%
  \BibitemOpen
  \bibfield  {author} {\bibinfo {author} {\bibfnamefont {Pascal~M.}\
  \bibnamefont {Vecsei}}, \bibinfo {author} {\bibfnamefont {Jose~L.}\
  \bibnamefont {Lado}}, \ and\ \bibinfo {author} {\bibfnamefont {Christian}\
  \bibnamefont {Flindt}},\ }\bibfield  {title} {\enquote {\bibinfo {title}
  {Lee-yang theory of the two-dimensional quantum ising model},}\ }\href
  {\doibase 10.1103/PhysRevB.106.054402} {\bibfield  {journal} {\bibinfo
  {journal} {Phys. Rev. B}\ }\textbf {\bibinfo {volume} {106}},\ \bibinfo
  {pages} {054402} (\bibinfo {year} {2022})}\BibitemShut {NoStop}%
\bibitem [{\citenamefont {Brange}\ \emph {et~al.}(2022)\citenamefont {Brange},
  \citenamefont {Peotta}, \citenamefont {Flindt},\ and\ \citenamefont
  {Ojanen}}]{Brange2022:Dynamical2Dquantum}%
  \BibitemOpen
  \bibfield  {author} {\bibinfo {author} {\bibfnamefont {Fredrik}\ \bibnamefont
  {Brange}}, \bibinfo {author} {\bibfnamefont {Sebastiano}\ \bibnamefont
  {Peotta}}, \bibinfo {author} {\bibfnamefont {Christian}\ \bibnamefont
  {Flindt}}, \ and\ \bibinfo {author} {\bibfnamefont {Teemu}\ \bibnamefont
  {Ojanen}},\ }\bibfield  {title} {\enquote {\bibinfo {title} {Dynamical
  quantum phase transitions in strongly correlated two-dimensional spin
  lattices following a quench},}\ }\href {\doibase
  10.1103/PhysRevResearch.4.033032} {\bibfield  {journal} {\bibinfo  {journal}
  {Phys. Rev. Res.}\ }\textbf {\bibinfo {volume} {4}},\ \bibinfo {pages}
  {033032} (\bibinfo {year} {2022})}\BibitemShut {NoStop}%
\bibitem [{\citenamefont {Vecsei}\ \emph {et~al.}(2023)\citenamefont {Vecsei},
  \citenamefont {Flindt},\ and\ \citenamefont {Lado}}]{vecsei2023leeyang}%
  \BibitemOpen
  \bibfield  {author} {\bibinfo {author} {\bibfnamefont {Pascal~M.}\
  \bibnamefont {Vecsei}}, \bibinfo {author} {\bibfnamefont {Christian}\
  \bibnamefont {Flindt}}, \ and\ \bibinfo {author} {\bibfnamefont {Jose~L.}\
  \bibnamefont {Lado}},\ }\href@noop {} {\enquote {\bibinfo {title} {Lee-yang
  theory of quantum phase transitions with neural network quantum states},}\ }
  (\bibinfo {year} {2023}),\ \Eprint {http://arxiv.org/abs/2301.09923}
  {arXiv:2301.09923 [cond-mat.str-el]} \BibitemShut {NoStop}%
\bibitem [{\citenamefont {Sandvik}(2010{\natexlab{a}})}]{Sandvik2010:CompStud}%
  \BibitemOpen
  \bibfield  {author} {\bibinfo {author} {\bibfnamefont {Anders~W.}\
  \bibnamefont {Sandvik}},\ }\bibfield  {title} {\enquote {\bibinfo {title}
  {Computational studies of quantum spin systems},}\ }\href {\doibase
  10.1063/1.3518900} {\bibfield  {journal} {\bibinfo  {journal} {AIP Conference
  Proceedings}\ }\textbf {\bibinfo {volume} {1297}},\ \bibinfo {pages}
  {135--338} (\bibinfo {year} {2010}{\natexlab{a}})},\ \Eprint
  {http://arxiv.org/abs/https://aip.scitation.org/doi/pdf/10.1063/1.3518900}
  {https://aip.scitation.org/doi/pdf/10.1063/1.3518900} \BibitemShut {NoStop}%
\bibitem [{\citenamefont {Wang}\ \emph {et~al.}(2006)\citenamefont {Wang},
  \citenamefont {Beach},\ and\ \citenamefont {Sandvik}}]{Wang2006:Bilayer}%
  \BibitemOpen
  \bibfield  {author} {\bibinfo {author} {\bibfnamefont {Ling}\ \bibnamefont
  {Wang}}, \bibinfo {author} {\bibfnamefont {K.~S.~D.}\ \bibnamefont {Beach}},
  \ and\ \bibinfo {author} {\bibfnamefont {Anders~W.}\ \bibnamefont
  {Sandvik}},\ }\bibfield  {title} {\enquote {\bibinfo {title} {High-precision
  finite-size scaling analysis of the quantum-critical point of {$S = 1/2$}
  heisenberg antiferromagnetic bilayers},}\ }\href {\doibase
  10.1103/PhysRevB.73.014431} {\bibfield  {journal} {\bibinfo  {journal} {Phys.
  Rev. B}\ }\textbf {\bibinfo {volume} {73}},\ \bibinfo {pages} {014431}
  (\bibinfo {year} {2006})}\BibitemShut {NoStop}%
\bibitem [{\citenamefont {Harris}(1970)}]{Harris1970:GenLeeYang}%
  \BibitemOpen
  \bibfield  {author} {\bibinfo {author} {\bibfnamefont {A.Brooks}\
  \bibnamefont {Harris}},\ }\bibfield  {title} {\enquote {\bibinfo {title}
  {Generalizations of the lee-yang theorem},}\ }\href {\doibase
  https://doi.org/10.1016/0375-9601(70)90708-5} {\bibfield  {journal} {\bibinfo
   {journal} {Physics Letters A}\ }\textbf {\bibinfo {volume} {33}},\ \bibinfo
  {pages} {161--162} (\bibinfo {year} {1970})}\BibitemShut {NoStop}%
\bibitem [{\citenamefont {Campostrini}\ \emph {et~al.}(2002)\citenamefont
  {Campostrini}, \citenamefont {Hasenbusch}, \citenamefont {Pelissetto},
  \citenamefont {Rossi},\ and\ \citenamefont
  {Vicari}}]{Campostrini2002:CritExpHeis}%
  \BibitemOpen
  \bibfield  {author} {\bibinfo {author} {\bibfnamefont {Massimo}\ \bibnamefont
  {Campostrini}}, \bibinfo {author} {\bibfnamefont {Martin}\ \bibnamefont
  {Hasenbusch}}, \bibinfo {author} {\bibfnamefont {Andrea}\ \bibnamefont
  {Pelissetto}}, \bibinfo {author} {\bibfnamefont {Paolo}\ \bibnamefont
  {Rossi}}, \ and\ \bibinfo {author} {\bibfnamefont {Ettore}\ \bibnamefont
  {Vicari}},\ }\bibfield  {title} {\enquote {\bibinfo {title} {Critical
  exponents and equation of state of the three-dimensional heisenberg
  universality class},}\ }\href {\doibase 10.1103/PhysRevB.65.144520}
  {\bibfield  {journal} {\bibinfo  {journal} {Phys. Rev. B}\ }\textbf {\bibinfo
  {volume} {65}},\ \bibinfo {pages} {144520} (\bibinfo {year}
  {2002})}\BibitemShut {NoStop}%
\bibitem [{\citenamefont {Itzykson}\ \emph {et~al.}(1983)\citenamefont
  {Itzykson}, \citenamefont {Pearson},\ and\ \citenamefont
  {Zuber}}]{Itzykson1983:DistributionZeros}%
  \BibitemOpen
  \bibfield  {author} {\bibinfo {author} {\bibfnamefont {C.}~\bibnamefont
  {Itzykson}}, \bibinfo {author} {\bibfnamefont {R.B.}\ \bibnamefont
  {Pearson}}, \ and\ \bibinfo {author} {\bibfnamefont {J.B.}\ \bibnamefont
  {Zuber}},\ }\bibfield  {title} {\enquote {\bibinfo {title} {Distribution of
  zeros in ising and gauge models},}\ }\href {\doibase
  https://doi.org/10.1016/0550-3213(83)90499-6} {\bibfield  {journal} {\bibinfo
   {journal} {Nuclear Physics B}\ }\textbf {\bibinfo {volume} {220}},\ \bibinfo
  {pages} {415--433} (\bibinfo {year} {1983})}\BibitemShut {NoStop}%
\bibitem [{\citenamefont {Gordillo-Guerrero}\ \emph {et~al.}(2013)\citenamefont
  {Gordillo-Guerrero}, \citenamefont {Kenna},\ and\ \citenamefont
  {Ruiz-Lorenzo}}]{Gordillo2013:LYZHeisenberg}%
  \BibitemOpen
  \bibfield  {author} {\bibinfo {author} {\bibfnamefont {A.}~\bibnamefont
  {Gordillo-Guerrero}}, \bibinfo {author} {\bibfnamefont {R.}~\bibnamefont
  {Kenna}}, \ and\ \bibinfo {author} {\bibfnamefont {J.~J.}\ \bibnamefont
  {Ruiz-Lorenzo}},\ }\bibfield  {title} {\enquote {\bibinfo {title} {Scaling
  behavior of the heisenberg model in three dimensions},}\ }\href {\doibase
  10.1103/PhysRevE.88.062117} {\bibfield  {journal} {\bibinfo  {journal} {Phys.
  Rev. E}\ }\textbf {\bibinfo {volume} {88}},\ \bibinfo {pages} {062117}
  (\bibinfo {year} {2013})}\BibitemShut {NoStop}%
\bibitem [{\citenamefont {Sandvik}(2007)}]{Sandvik2007:JQ}%
  \BibitemOpen
  \bibfield  {author} {\bibinfo {author} {\bibfnamefont {Anders~W.}\
  \bibnamefont {Sandvik}},\ }\bibfield  {title} {\enquote {\bibinfo {title}
  {Evidence for deconfined quantum criticality in a two-dimensional heisenberg
  model with four-spin interactions},}\ }\href {\doibase
  10.1103/PhysRevLett.98.227202} {\bibfield  {journal} {\bibinfo  {journal}
  {Phys. Rev. Lett.}\ }\textbf {\bibinfo {volume} {98}},\ \bibinfo {pages}
  {227202} (\bibinfo {year} {2007})}\BibitemShut {NoStop}%
\bibitem [{\citenamefont {Senthil}\ \emph
  {et~al.}(2004{\natexlab{a}})\citenamefont {Senthil}, \citenamefont
  {Vishwanath}, \citenamefont {Balents}, \citenamefont {Sachdev},\ and\
  \citenamefont {Fisher}}]{Senthil2004:DQCPs}%
  \BibitemOpen
  \bibfield  {author} {\bibinfo {author} {\bibfnamefont {T.}~\bibnamefont
  {Senthil}}, \bibinfo {author} {\bibfnamefont {Ashvin}\ \bibnamefont
  {Vishwanath}}, \bibinfo {author} {\bibfnamefont {Leon}\ \bibnamefont
  {Balents}}, \bibinfo {author} {\bibfnamefont {Subir}\ \bibnamefont
  {Sachdev}}, \ and\ \bibinfo {author} {\bibfnamefont {Matthew P.~A.}\
  \bibnamefont {Fisher}},\ }\bibfield  {title} {\enquote {\bibinfo {title}
  {Deconfined quantum critical points},}\ }\href {\doibase
  10.1126/science.1091806} {\bibfield  {journal} {\bibinfo  {journal}
  {Science}\ }\textbf {\bibinfo {volume} {303}},\ \bibinfo {pages} {1490--1494}
  (\bibinfo {year} {2004}{\natexlab{a}})},\ \Eprint
  {http://arxiv.org/abs/https://www.science.org/doi/pdf/10.1126/science.1091806}
  {https://www.science.org/doi/pdf/10.1126/science.1091806} \BibitemShut
  {NoStop}%
\bibitem [{\citenamefont {Senthil}\ \emph
  {et~al.}(2004{\natexlab{b}})\citenamefont {Senthil}, \citenamefont {Balents},
  \citenamefont {Sachdev}, \citenamefont {Vishwanath},\ and\ \citenamefont
  {Fisher}}]{Senthil2004:BeyondLGW}%
  \BibitemOpen
  \bibfield  {author} {\bibinfo {author} {\bibfnamefont {T.}~\bibnamefont
  {Senthil}}, \bibinfo {author} {\bibfnamefont {Leon}\ \bibnamefont {Balents}},
  \bibinfo {author} {\bibfnamefont {Subir}\ \bibnamefont {Sachdev}}, \bibinfo
  {author} {\bibfnamefont {Ashvin}\ \bibnamefont {Vishwanath}}, \ and\ \bibinfo
  {author} {\bibfnamefont {Matthew P.~A.}\ \bibnamefont {Fisher}},\ }\bibfield
  {title} {\enquote {\bibinfo {title} {Quantum criticality beyond the
  landau-ginzburg-wilson paradigm},}\ }\href {\doibase
  10.1103/PhysRevB.70.144407} {\bibfield  {journal} {\bibinfo  {journal} {Phys.
  Rev. B}\ }\textbf {\bibinfo {volume} {70}},\ \bibinfo {pages} {144407}
  (\bibinfo {year} {2004}{\natexlab{b}})}\BibitemShut {NoStop}%
\bibitem [{\citenamefont {Lou}\ \emph {et~al.}(2009)\citenamefont {Lou},
  \citenamefont {Sandvik},\ and\ \citenamefont
  {Kawashima}}]{Lou2009:AFtoVBSsun}%
  \BibitemOpen
  \bibfield  {author} {\bibinfo {author} {\bibfnamefont {Jie}\ \bibnamefont
  {Lou}}, \bibinfo {author} {\bibfnamefont {Anders~W.}\ \bibnamefont
  {Sandvik}}, \ and\ \bibinfo {author} {\bibfnamefont {Naoki}\ \bibnamefont
  {Kawashima}},\ }\bibfield  {title} {\enquote {\bibinfo {title}
  {Antiferromagnetic to valence-bond-solid transitions in two-dimensional
  $\text{SU}(n)$ heisenberg models with multispin interactions},}\ }\href
  {\doibase 10.1103/PhysRevB.80.180414} {\bibfield  {journal} {\bibinfo
  {journal} {Phys. Rev. B}\ }\textbf {\bibinfo {volume} {80}},\ \bibinfo
  {pages} {180414} (\bibinfo {year} {2009})}\BibitemShut {NoStop}%
\bibitem [{\citenamefont
  {Sandvik}(2010{\natexlab{b}})}]{Sandvik2010:ContinuousJQlog}%
  \BibitemOpen
  \bibfield  {author} {\bibinfo {author} {\bibfnamefont {Anders~W.}\
  \bibnamefont {Sandvik}},\ }\bibfield  {title} {\enquote {\bibinfo {title}
  {Continuous quantum phase transition between an antiferromagnet and a
  valence-bond solid in two dimensions: Evidence for logarithmic corrections to
  scaling},}\ }\href {\doibase 10.1103/PhysRevLett.104.177201} {\bibfield
  {journal} {\bibinfo  {journal} {Phys. Rev. Lett.}\ }\textbf {\bibinfo
  {volume} {104}},\ \bibinfo {pages} {177201} (\bibinfo {year}
  {2010}{\natexlab{b}})}\BibitemShut {NoStop}%
\bibitem [{\citenamefont {Harada}\ \emph {et~al.}(2013)\citenamefont {Harada},
  \citenamefont {Suzuki}, \citenamefont {Okubo}, \citenamefont {Matsuo},
  \citenamefont {Lou}, \citenamefont {Watanabe}, \citenamefont {Todo},\ and\
  \citenamefont {Kawashima}}]{Harada2013:DQCPsmallN}%
  \BibitemOpen
  \bibfield  {author} {\bibinfo {author} {\bibfnamefont {Kenji}\ \bibnamefont
  {Harada}}, \bibinfo {author} {\bibfnamefont {Takafumi}\ \bibnamefont
  {Suzuki}}, \bibinfo {author} {\bibfnamefont {Tsuyoshi}\ \bibnamefont
  {Okubo}}, \bibinfo {author} {\bibfnamefont {Haruhiko}\ \bibnamefont
  {Matsuo}}, \bibinfo {author} {\bibfnamefont {Jie}\ \bibnamefont {Lou}},
  \bibinfo {author} {\bibfnamefont {Hiroshi}\ \bibnamefont {Watanabe}},
  \bibinfo {author} {\bibfnamefont {Synge}\ \bibnamefont {Todo}}, \ and\
  \bibinfo {author} {\bibfnamefont {Naoki}\ \bibnamefont {Kawashima}},\
  }\bibfield  {title} {\enquote {\bibinfo {title} {Possibility of deconfined
  criticality in su($n$) heisenberg models at small $n$},}\ }\href {\doibase
  10.1103/PhysRevB.88.220408} {\bibfield  {journal} {\bibinfo  {journal} {Phys.
  Rev. B}\ }\textbf {\bibinfo {volume} {88}},\ \bibinfo {pages} {220408}
  (\bibinfo {year} {2013})}\BibitemShut {NoStop}%
\bibitem [{\citenamefont {Block}\ \emph {et~al.}(2013)\citenamefont {Block},
  \citenamefont {Melko},\ and\ \citenamefont {Kaul}}]{Block2013:Fate}%
  \BibitemOpen
  \bibfield  {author} {\bibinfo {author} {\bibfnamefont {Matthew~S.}\
  \bibnamefont {Block}}, \bibinfo {author} {\bibfnamefont {Roger~G.}\
  \bibnamefont {Melko}}, \ and\ \bibinfo {author} {\bibfnamefont {Ribhu~K.}\
  \bibnamefont {Kaul}},\ }\bibfield  {title} {\enquote {\bibinfo {title} {Fate
  of $\mathbb{C}{\mathbb{p}}^{N\ensuremath{-}1}$ fixed points with $q$
  monopoles},}\ }\href {\doibase 10.1103/PhysRevLett.111.137202} {\bibfield
  {journal} {\bibinfo  {journal} {Phys. Rev. Lett.}\ }\textbf {\bibinfo
  {volume} {111}},\ \bibinfo {pages} {137202} (\bibinfo {year}
  {2013})}\BibitemShut {NoStop}%
\bibitem [{\citenamefont {Shao}\ \emph {et~al.}(2016)\citenamefont {Shao},
  \citenamefont {Guo},\ and\ \citenamefont {Sandvik}}]{Shao2016:TwoLength}%
  \BibitemOpen
  \bibfield  {author} {\bibinfo {author} {\bibfnamefont {Hui}\ \bibnamefont
  {Shao}}, \bibinfo {author} {\bibfnamefont {Wenan}\ \bibnamefont {Guo}}, \
  and\ \bibinfo {author} {\bibfnamefont {Anders~W.}\ \bibnamefont {Sandvik}},\
  }\bibfield  {title} {\enquote {\bibinfo {title} {Quantum criticality with two
  length scales},}\ }\href {\doibase 10.1126/science.aad5007} {\bibfield
  {journal} {\bibinfo  {journal} {Science}\ }\textbf {\bibinfo {volume}
  {352}},\ \bibinfo {pages} {213--216} (\bibinfo {year} {2016})},\ \Eprint
  {http://arxiv.org/abs/https://www.science.org/doi/pdf/10.1126/science.aad5007}
  {https://www.science.org/doi/pdf/10.1126/science.aad5007} \BibitemShut
  {NoStop}%
\bibitem [{\citenamefont {Sandvik}\ and\ \citenamefont
  {Zhao}(2020)}]{Sandvik2020:Consistent}%
  \BibitemOpen
  \bibfield  {author} {\bibinfo {author} {\bibfnamefont {Anders~W.}\
  \bibnamefont {Sandvik}}\ and\ \bibinfo {author} {\bibfnamefont {Bowen}\
  \bibnamefont {Zhao}},\ }\bibfield  {title} {\enquote {\bibinfo {title}
  {Consistent scaling exponents at the deconfined quantum-critical point*},}\
  }\href {\doibase 10.1088/0256-307X/37/5/057502} {\bibfield  {journal}
  {\bibinfo  {journal} {Chinese Physics Letters}\ }\textbf {\bibinfo {volume}
  {37}},\ \bibinfo {pages} {057502} (\bibinfo {year} {2020})}\BibitemShut
  {NoStop}%
\bibitem [{\citenamefont {Zhao}\ \emph {et~al.}(2020)\citenamefont {Zhao},
  \citenamefont {Takahashi},\ and\ \citenamefont {Sandvik}}]{Zhao2020:Helical}%
  \BibitemOpen
  \bibfield  {author} {\bibinfo {author} {\bibfnamefont {Bowen}\ \bibnamefont
  {Zhao}}, \bibinfo {author} {\bibfnamefont {Jun}\ \bibnamefont {Takahashi}}, \
  and\ \bibinfo {author} {\bibfnamefont {Anders~W.}\ \bibnamefont {Sandvik}},\
  }\bibfield  {title} {\enquote {\bibinfo {title} {Multicritical deconfined
  quantum criticality and lifshitz point of a helical valence-bond phase},}\
  }\href {\doibase 10.1103/PhysRevLett.125.257204} {\bibfield  {journal}
  {\bibinfo  {journal} {Phys. Rev. Lett.}\ }\textbf {\bibinfo {volume} {125}},\
  \bibinfo {pages} {257204} (\bibinfo {year} {2020})}\BibitemShut {NoStop}%
\bibitem [{\citenamefont {Kuklov}\ \emph {et~al.}(2008)\citenamefont {Kuklov},
  \citenamefont {Matsumoto}, \citenamefont {Prokof'ev}, \citenamefont
  {Svistunov},\ and\ \citenamefont {Troyer}}]{Kuklov2008:DCPfirstorder}%
  \BibitemOpen
  \bibfield  {author} {\bibinfo {author} {\bibfnamefont {A.~B.}\ \bibnamefont
  {Kuklov}}, \bibinfo {author} {\bibfnamefont {M.}~\bibnamefont {Matsumoto}},
  \bibinfo {author} {\bibfnamefont {N.~V.}\ \bibnamefont {Prokof'ev}}, \bibinfo
  {author} {\bibfnamefont {B.~V.}\ \bibnamefont {Svistunov}}, \ and\ \bibinfo
  {author} {\bibfnamefont {M.}~\bibnamefont {Troyer}},\ }\bibfield  {title}
  {\enquote {\bibinfo {title} {Deconfined criticality: Generic first-order
  transition in the su(2) symmetry case},}\ }\href {\doibase
  10.1103/PhysRevLett.101.050405} {\bibfield  {journal} {\bibinfo  {journal}
  {Phys. Rev. Lett.}\ }\textbf {\bibinfo {volume} {101}},\ \bibinfo {pages}
  {050405} (\bibinfo {year} {2008})}\BibitemShut {NoStop}%
\bibitem [{\citenamefont {Jiang}\ \emph {et~al.}(2008)\citenamefont {Jiang},
  \citenamefont {Nyfeler}, \citenamefont {Chandrasekharan},\ and\ \citenamefont
  {Wiese}}]{Jiang2008:FirstOrder}%
  \BibitemOpen
  \bibfield  {author} {\bibinfo {author} {\bibfnamefont {F-J}\ \bibnamefont
  {Jiang}}, \bibinfo {author} {\bibfnamefont {M}~\bibnamefont {Nyfeler}},
  \bibinfo {author} {\bibfnamefont {S}~\bibnamefont {Chandrasekharan}}, \ and\
  \bibinfo {author} {\bibfnamefont {U-J}\ \bibnamefont {Wiese}},\ }\bibfield
  {title} {\enquote {\bibinfo {title} {From an antiferromagnet to a valence
  bond solid: evidence for a first-order phase transition},}\ }\href {\doibase
  10.1088/1742-5468/2008/02/P02009} {\bibfield  {journal} {\bibinfo  {journal}
  {Journal of Statistical Mechanics: Theory and Experiment}\ }\textbf {\bibinfo
  {volume} {2008}},\ \bibinfo {pages} {P02009} (\bibinfo {year}
  {2008})}\BibitemShut {NoStop}%
\bibitem [{\citenamefont {Chen}\ \emph {et~al.}(2013)\citenamefont {Chen},
  \citenamefont {Huang}, \citenamefont {Deng}, \citenamefont {Kuklov},
  \citenamefont {Prokof'ev},\ and\ \citenamefont
  {Svistunov}}]{Chen2013:DCPflow}%
  \BibitemOpen
  \bibfield  {author} {\bibinfo {author} {\bibfnamefont {Kun}\ \bibnamefont
  {Chen}}, \bibinfo {author} {\bibfnamefont {Yuan}\ \bibnamefont {Huang}},
  \bibinfo {author} {\bibfnamefont {Youjin}\ \bibnamefont {Deng}}, \bibinfo
  {author} {\bibfnamefont {A.~B.}\ \bibnamefont {Kuklov}}, \bibinfo {author}
  {\bibfnamefont {N.~V.}\ \bibnamefont {Prokof'ev}}, \ and\ \bibinfo {author}
  {\bibfnamefont {B.~V.}\ \bibnamefont {Svistunov}},\ }\bibfield  {title}
  {\enquote {\bibinfo {title} {Deconfined criticality flow in the heisenberg
  model with ring-exchange interactions},}\ }\href {\doibase
  10.1103/PhysRevLett.110.185701} {\bibfield  {journal} {\bibinfo  {journal}
  {Phys. Rev. Lett.}\ }\textbf {\bibinfo {volume} {110}},\ \bibinfo {pages}
  {185701} (\bibinfo {year} {2013})}\BibitemShut {NoStop}%
\bibitem [{\citenamefont {D'Emidio}\ \emph {et~al.}(2021)\citenamefont
  {D'Emidio}, \citenamefont {Eberharter},\ and\ \citenamefont
  {Läuchli}}]{demidio2021diagnosing}%
  \BibitemOpen
  \bibfield  {author} {\bibinfo {author} {\bibfnamefont {Jonathan}\
  \bibnamefont {D'Emidio}}, \bibinfo {author} {\bibfnamefont {Alexander~A.}\
  \bibnamefont {Eberharter}}, \ and\ \bibinfo {author} {\bibfnamefont
  {Andreas~M.}\ \bibnamefont {Läuchli}},\ }\href@noop {} {\enquote {\bibinfo
  {title} {Diagnosing weakly first-order phase transitions by coupling to order
  parameters},}\ } (\bibinfo {year} {2021}),\ \Eprint
  {http://arxiv.org/abs/2106.15462} {arXiv:2106.15462 [cond-mat.str-el]}
  \BibitemShut {NoStop}%
\bibitem [{\citenamefont {Zhao}\ \emph {et~al.}(2022)\citenamefont {Zhao},
  \citenamefont {Wang}, \citenamefont {Yan}, \citenamefont {Cheng},\ and\
  \citenamefont {Meng}}]{Zhao2022:EEatDQC}%
  \BibitemOpen
  \bibfield  {author} {\bibinfo {author} {\bibfnamefont {Jiarui}\ \bibnamefont
  {Zhao}}, \bibinfo {author} {\bibfnamefont {Yan-Cheng}\ \bibnamefont {Wang}},
  \bibinfo {author} {\bibfnamefont {Zheng}\ \bibnamefont {Yan}}, \bibinfo
  {author} {\bibfnamefont {Meng}\ \bibnamefont {Cheng}}, \ and\ \bibinfo
  {author} {\bibfnamefont {Zi~Yang}\ \bibnamefont {Meng}},\ }\bibfield  {title}
  {\enquote {\bibinfo {title} {Scaling of entanglement entropy at deconfined
  quantum criticality},}\ }\href {\doibase 10.1103/PhysRevLett.128.010601}
  {\bibfield  {journal} {\bibinfo  {journal} {Phys. Rev. Lett.}\ }\textbf
  {\bibinfo {volume} {128}},\ \bibinfo {pages} {010601} (\bibinfo {year}
  {2022})}\BibitemShut {NoStop}%
\bibitem [{\citenamefont {Song}\ \emph {et~al.}(2023)\citenamefont {Song},
  \citenamefont {Zhao}, \citenamefont {Janssen}, \citenamefont {Scherer},\ and\
  \citenamefont {Meng}}]{song2023deconfined}%
  \BibitemOpen
  \bibfield  {author} {\bibinfo {author} {\bibfnamefont {Menghan}\ \bibnamefont
  {Song}}, \bibinfo {author} {\bibfnamefont {Jiarui}\ \bibnamefont {Zhao}},
  \bibinfo {author} {\bibfnamefont {Lukas}\ \bibnamefont {Janssen}}, \bibinfo
  {author} {\bibfnamefont {Michael~M.}\ \bibnamefont {Scherer}}, \ and\
  \bibinfo {author} {\bibfnamefont {Zi~Yang}\ \bibnamefont {Meng}},\
  }\href@noop {} {\enquote {\bibinfo {title} {Deconfined quantum criticality
  lost},}\ } (\bibinfo {year} {2023}),\ \Eprint
  {http://arxiv.org/abs/2307.02547} {arXiv:2307.02547 [cond-mat.str-el]}
  \BibitemShut {NoStop}%
\bibitem [{\citenamefont {Nahum}\ \emph {et~al.}(2015)\citenamefont {Nahum},
  \citenamefont {Chalker}, \citenamefont {Serna}, \citenamefont {Ortu\~no},\
  and\ \citenamefont {Somoza}}]{Nahum2015:DQCloop}%
  \BibitemOpen
  \bibfield  {author} {\bibinfo {author} {\bibfnamefont {Adam}\ \bibnamefont
  {Nahum}}, \bibinfo {author} {\bibfnamefont {J.~T.}\ \bibnamefont {Chalker}},
  \bibinfo {author} {\bibfnamefont {P.}~\bibnamefont {Serna}}, \bibinfo
  {author} {\bibfnamefont {M.}~\bibnamefont {Ortu\~no}}, \ and\ \bibinfo
  {author} {\bibfnamefont {A.~M.}\ \bibnamefont {Somoza}},\ }\bibfield  {title}
  {\enquote {\bibinfo {title} {Deconfined quantum criticality, scaling
  violations, and classical loop models},}\ }\href {\doibase
  10.1103/PhysRevX.5.041048} {\bibfield  {journal} {\bibinfo  {journal} {Phys.
  Rev. X}\ }\textbf {\bibinfo {volume} {5}},\ \bibinfo {pages} {041048}
  (\bibinfo {year} {2015})}\BibitemShut {NoStop}%
\bibitem [{\citenamefont {Gorbenko}\ \emph {et~al.}(2018)\citenamefont
  {Gorbenko}, \citenamefont {Rychkov},\ and\ \citenamefont
  {Zan}}]{Gorbenko2018:Walking}%
  \BibitemOpen
  \bibfield  {author} {\bibinfo {author} {\bibfnamefont {Victor}\ \bibnamefont
  {Gorbenko}}, \bibinfo {author} {\bibfnamefont {Slava}\ \bibnamefont
  {Rychkov}}, \ and\ \bibinfo {author} {\bibfnamefont {Bernardo}\ \bibnamefont
  {Zan}},\ }\bibfield  {title} {\enquote {\bibinfo {title} {Walking, weak
  first-order transitions, and complex cfts},}\ }\href {\doibase
  10.1007/JHEP10(2018)108} {\bibfield  {journal} {\bibinfo  {journal} {Journal
  of High Energy Physics}\ }\textbf {\bibinfo {volume} {2018}},\ \bibinfo
  {pages} {108} (\bibinfo {year} {2018})}\BibitemShut {NoStop}%
\bibitem [{\citenamefont {Ma}\ and\ \citenamefont
  {Wang}(2020)}]{Ma2020:Pseudo}%
  \BibitemOpen
  \bibfield  {author} {\bibinfo {author} {\bibfnamefont {Ruochen}\ \bibnamefont
  {Ma}}\ and\ \bibinfo {author} {\bibfnamefont {Chong}\ \bibnamefont {Wang}},\
  }\bibfield  {title} {\enquote {\bibinfo {title} {Theory of deconfined
  pseudocriticality},}\ }\href {\doibase 10.1103/PhysRevB.102.020407}
  {\bibfield  {journal} {\bibinfo  {journal} {Phys. Rev. B}\ }\textbf {\bibinfo
  {volume} {102}},\ \bibinfo {pages} {020407} (\bibinfo {year}
  {2020})}\BibitemShut {NoStop}%
\bibitem [{\citenamefont {Nahum}(2020)}]{Nahum2020:Quasi}%
  \BibitemOpen
  \bibfield  {author} {\bibinfo {author} {\bibfnamefont {Adam}\ \bibnamefont
  {Nahum}},\ }\bibfield  {title} {\enquote {\bibinfo {title} {Note on
  wess-zumino-witten models and quasiuniversality in $2+1$ dimensions},}\
  }\href {\doibase 10.1103/PhysRevB.102.201116} {\bibfield  {journal} {\bibinfo
   {journal} {Phys. Rev. B}\ }\textbf {\bibinfo {volume} {102}},\ \bibinfo
  {pages} {201116} (\bibinfo {year} {2020})}\BibitemShut {NoStop}%
\end{thebibliography}%

\clearpage
%%\widetext

\begin{figure}[!t]
\centerline{\includegraphics[angle=0,width=1.0\columnwidth]{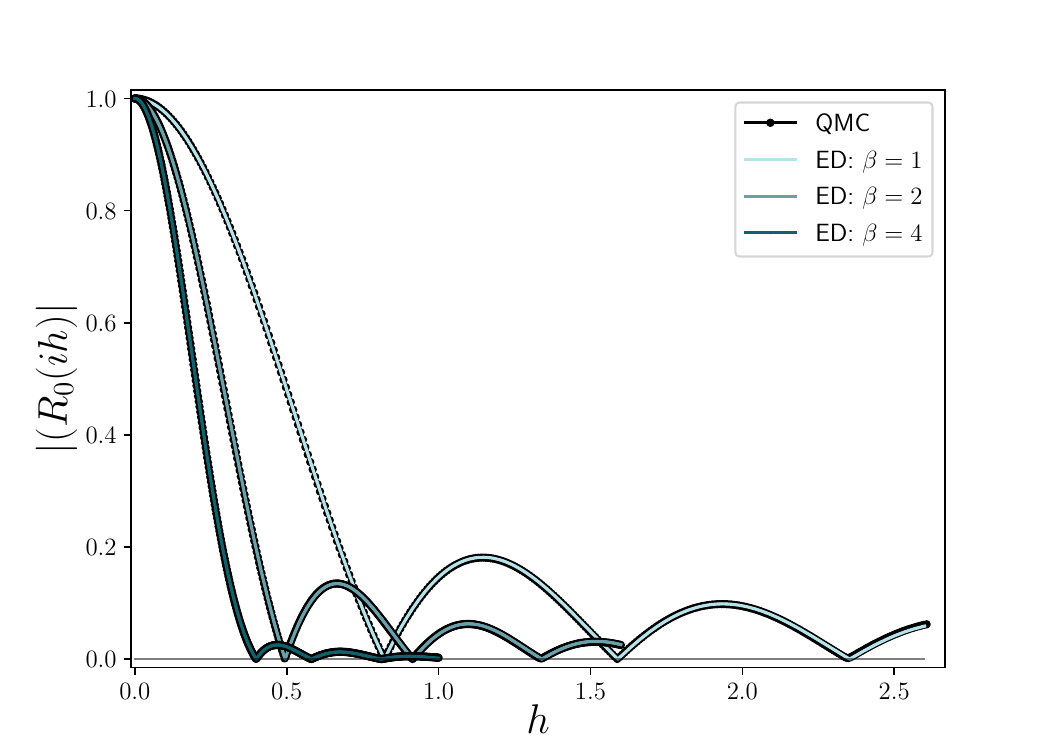}}
\caption{A comparison of data obtained with the QMC method versus exact diagonalization.  The system is an $L=2$ Heisenberg bilayer with $J=1, J_{\perp}=2.52205$ with three different values of $\beta$.  The ratio $|R_0(h)|$ is plotted as a function of the N\'eel field $h$ along the imaginary axis.}
\label{fig:QMCvED}
\end{figure}

\begin{figure}[!t]
\centerline{\includegraphics[angle=0,width=1.2\columnwidth]{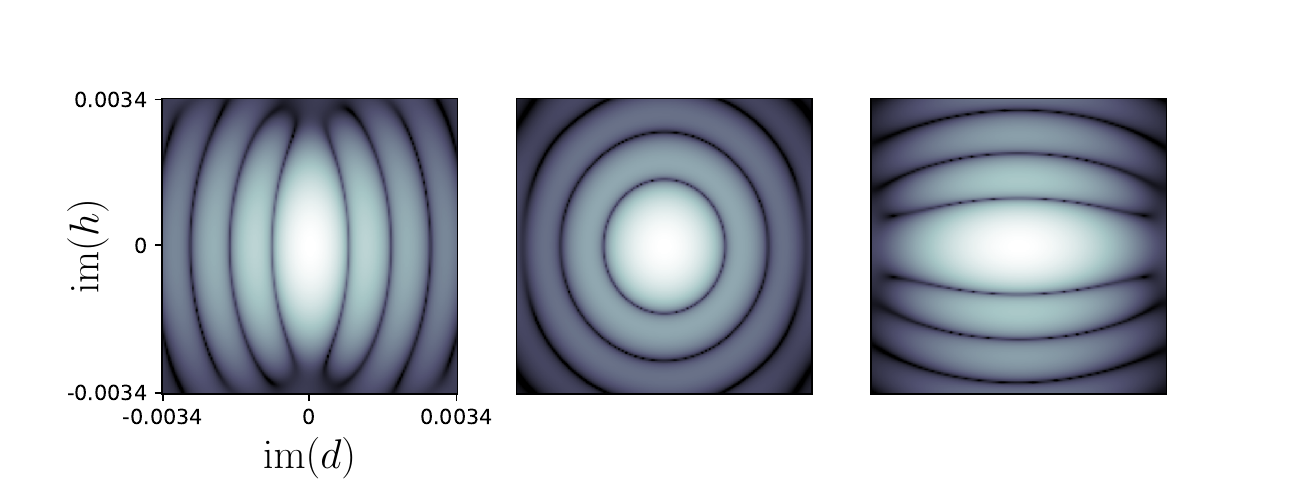}}
\caption{A density plot of $-\ln(|R_0(h,d)|)$ for the $J$-$Q$ model at the critical coupling with $L=40$ at three different values of the coupling: $J/Q=0.015, 0.045, 0.085$ from left to right.}
\label{fig:JQextra}
\end{figure}

\section{Supplemental material}

\subsection{QMC versus exact diagonalization}
Here we demonstrate the validity of Eqn. (\ref{eq:zrath}) by comparing against exact results for a small system size.  Fig. (\ref{fig:QMCvED}) shows $|R_0(h)|$ computed with a purely imaginary N\'eel field on a $2 \times 2$ Heisenberg bilayer system compared against exact diagonalization of the whole spectrum.  We make the comparison for three different values of $\beta$, where in each case the first three Lee-Yang zeros are visible.  The QMC values match the exact results perfectly within the statistical uncertainty.

\subsection{Data away from deconfined critical point}
Here we would like to show what happens to the nearly circular distribution of zeros, with combined imaginary N\'eel and VBS fields, away from the critical point in the $J$-$Q$ model.  In Fig. (\ref{fig:JQextra}) we show density plots (similar to Fig. (\ref{fig:jqmap})) for three different values of the coupling: $J/Q=0.015, 0.045, 0.085$ from left to right for system size $L=40$.  Here we see on the left, which is on the VBS side of the transition, the N\'eel zeros get pushed away from the origin, relative to their values at the transition given, and in fact are not clearly detectable on this plot.  A similar story goes for the right panel.  While we believe our data to be statistically converged for the field values shown, which would indicate a potential rearrangement of the lines of zeros away from the critical point, we cannot rule out the possibility of under sampling at the larger field values.

\end{document}